\documentclass[iop,apj,twocolappendix,numberedappendix,revtex4]{emulateapj}	

\usepackage[breaklinks,colorlinks,citecolor=blue]{hyperref}		
\usepackage{color}      
\usepackage{amsmath}
\usepackage[thickspace,squaren]{SIunits} 

\def\mathbi#1{\textbf{\em #1}}
\newcommand{\erf}{\ensuremath{\text{erf}}{\vphantom{\Bigl(\Bigr)}}} 

\renewcommand{\vec}[1]{\ensuremath{\mathbi{#1}}} 
\newcommand{\gvec}[1]{\ensuremath{\mbox{\boldmath$ #1 $}}} 
\newcommand{\uvec}[1]{\ensuremath{\hat{\mathbi{#1}}}} 
\newcommand{\cprod}{\ensuremath{\pmb{\times}}} 
\newcommand{\abs}[1]{\left| #1 \right|} 
\newcommand{\pd}[2]{\frac{\partial #1}{\partial #2}} 
\newcommand{\pdd}[2]{\frac{\partial^2 #1}{\partial #2^2}} 
\newcommand{\curl}[1]{\gvec{\nabla} \cprod #1} 
\providecommand{\abs}[1]{\lvert#1\rvert}

\newcommand{\ps}{\reciprocal\second}
\newcommand{\mps}{\metre\:\ps}      
\newcommand{\mmps}{\metre\squared\:\ps}      
\newcommand{\cmmps}{\centi\mmps}      
\newcommand{\kmmps}{\kilo\mmps}      
\newcommand{\radps}{\radian\:\ps}      
\newcommand{\gauss}{\text{G}}
\newcommand{\maxwell}{\text{Mx}}
\newcommand{\years}{\text{years}}
\renewcommand{\phm}{\phantom}
\newcommand{\tblmk}{\tablenotemark}
\newcommand{\sstk}{\substack}
\newcommand{\8}{\infty}

\newcommand{\BMR}{\text{BMR}}    
\newcommand{\GA}{\text{GA}}      
\newcommand{\PDF}{\text{PDF}}    
\newcommand{\SFT}{\text{SFT}}    
\newcommand{\WS}{\text{WS}}      

\shorttitle{$2\times2$D Babcock--Leighton dynamo. I. Surface}
\shortauthors{A. Lemerle, P. Charbonneau, A. Carignan-Dugas}

\begin{document}


\submitted{Draft version 2015 August 24; see ApJ published version: 
\href{http://dx.doi.org/10.1088/0004-637X/810/1/78}
{http://dx.doi.org/10.1088/0004-637X/810/1/78}}

\title{A coupled $2\times2$D Babcock--Leighton solar dynamo model.\\
I. Surface magnetic flux evolution}


\author{Alexandre Lemerle\altaffilmark{1,2}}
\author{Paul Charbonneau\altaffilmark{1}}
\author{Arnaud Carignan-Dugas\altaffilmark{1}}

\altaffiltext{1}{D\'epartement de physique, Universit\'e de Montr\'eal, 
2900 boul. \'Edouard-Montpetit, Montr\'eal, QC, H3T 1J4, Canada;
lemerle@astro.umontreal.ca, paulchar@astro.umontreal.ca}
\altaffiltext{2}{Coll\`ege de Bois-de-Boulogne,
10555 av. Bois-de-Boulogne, Montr\'eal, QC, H4N 1L4, Canada.}


\begin{abstract}
The need for reliable predictions of the solar activity cycle motivates 
the development of dynamo models incorporating a representation 
of surface processes sufficiently detailed to allow assimilation of 
magnetographic data.
In this series of papers we present one such dynamo model, and document
its behavior and properties.
This first paper focuses on one of the model's key components, namely 
surface magnetic flux evolution.
Using a genetic algorithm, we obtain best-fit parameters of the 
transport model by least-squares minimization of the differences between
the associated synthetic synoptic magnetogram and real magnetographic 
data for activity cycle 21.
Our fitting procedure also returns Monte Carlo-like error estimates.
We show that the range of acceptable surface meridional flow profiles is 
in good agreement with Doppler measurements, even though the latter are 
not used in the fitting process.
Using a synthetic database of bipolar magnetic region (BMR) emergences 
reproducing the 
statistical properties of observed emergences, we also ascertain the 
sensitivity of global cycle properties, such as the strength of the 
dipole moment and timing of polarity reversal, to distinct 
realizations of BMR emergence, and on this basis argue that 
this stochasticity represents a primary source of uncertainty for 
predicting solar cycle characteristics.
\end{abstract}


\section{Introduction}

The Sun's magnetic field is generated by a magnetohydrodynamical 
induction process, or a combination of processes, taking place primarily
in the solar convection zone.
On small spatial scales, convection is believed to continuously process 
and replenish the photospheric magnetic field, through a local dynamo 
mechanism that is statistically stationary and does not produce net 
signed flux.
At the other extreme, magnetic fields developing on spatial scales 
commensurate with the solar radius show a strong degree of axisymmetry 
and a well-defined dipole moment, and undergo polarity reversals on a 
regular cadence of approximately eleven years 
(see review by \citealt{Hathaway2010}).

Cowling's theorem dictates that such an axisymmetric large-scale
magnetic field cannot be sustained by purely axisymmetric flows.
Convective turbulence represents an ideal energy reservoir for the 
required dynamo action, provided the Coriolis force can break the
mirror symmetry that would otherwise prevail.
This process can be quantified using mean-field electrodynamics, leading
to the so-called $\alpha$-effect, an electromotive force proportional to
the mean magnetic fields 
(for a recent review see, e.g., \citealt{Charbonneau2014}).
That such a turbulent dynamo, acting in conjunction with differential
rotation, can lead to the production of large-scale magnetic fields
undergoing polarity reversals has
been confirmed by both laboratory experiments
\citep{Lathrop2011,Cooper2014,Zimmerman2014} and
global magnetohydrodynamical (MHD) numerical simulations of solar 
convection
(see \citealt[\S~3.2]{Charbonneau2014}, and references therein).

The Coriolis force also acts on the flows developing along the axis of 
buoyantly rising toroidal magnetic flux ropes, believed to be generated 
near the base of the solar convection zone, and eventually piercing the 
photosphere in the form of bipolar magnetic regions (hereafter {\BMR}s; 
see \citealt{Fan2009} for a review).
This rotational influence produces the observed systematic east--west 
tilt characterizing large {\BMR}s, as embodied in Joy's Law.
Associated with this tilt is a net dipole moment so that, effectively, a
poloidal magnetic component is being produced from a pre-existing 
toroidal component.
Here again it is the Coriolis force that ultimately breaks the 
axisymmetry of the initially purely toroidal flux rope, so the process 
is akin to a large-scale version of the $\alpha$-effect. With shearing 
by differential rotation producing a toroidal magnetic component from a 
pre-existing poloidal component, the dynamo loop can be closed.
This forms the basis of the Babcock--Leighton dynamo models
\citep{Babcock1961,Leighton1969}, which have undergone a strong revival 
in the past two decades and are now considered a leading explanatory 
framework for the solar magnetic cycle
(for a recent review, see, e.g., \citealt{Karak2014}).

In such models,
the transport and accumulation in polar regions of the magnetic flux 
liberated at low latitudes by the decay of a {\BMR} is what sets 
the magnitude of the resulting dipole moment and the timing of its 
reversal \citep{Wang1989,Wang1991}.
The cross-equatorial diffusive annihilation of magnetic flux associated 
with the leading members of tilted {\BMR}s is ultimately what allows the 
build-up of a net hemispheric signed flux
(see \citealt{Cameron2013,Cameron2014}, and references therein).
Indeed, only a small fraction of emerging magnetic flux eventually makes
it to the poles; the magnetic flux in the polar cap at sunspot minimum,
$\simeq\unit{10^{22}}{\maxwell}$, is about the same as the unsigned flux
in a single, large BMR, and the net axial dipole moment of all BMRs
emerging
during a typical cycle is a few times the dipole moment required for 
polarity inversion \citep{Wang1989-0}.
Consequently, one large {\BMR} emerging very close to the equator with a 
significant tilt can have a strong impact on the magnitude of the dipole 
moment building up in the descending phase of the cycle, and thus on the 
amplitude of the subsequent cycle (see, e.g., \citealt{Jiang2014}).

In this series of papers we present a novel Babcock--Leighton model of 
the solar cycle based on the coupling of a surface flux transport 
({\SFT}) simulation with a mean-field-like interior dynamo model.
We henceforth refer to this hybrid as a ``$2\times2$D model'', as it 
couples
a two-dimensional simulation on a spherical surface ($\theta,\phi$) to
a two-dimensional simulation on a meridional plane ($r,\theta$), each
simulation providing the source term required by the other.

In the present paper we focus on the SFT component of the model.
SFT has been extensively studied in the past decades, starting with the work
of \citet{Leighton1964} up to recent attempts to reproduce the details
of modern magnetograms (see reviews by \citealt{Sheeley2005}, 
\citealt{Mackay2012}, and \citealt{Jiang2014r}).
The model's behavior relative to emergence and model characteristics are 
fairly well understood (see, e.g., \citealt{Baumann2004}).
In particular, observed magnetographic features, such as poleward flux 
strips (``surges''), require a delicate balance between 
meridional circulation and the surface effective diffusion rate 
(see, e.g. \citealt{Wang1989-2}), ultimately driven by the dispersive random 
walk taking place at the supergranular scale.
Yet, due to limitations in the measurement of these two processes,
their detailed parameterization remains, even today, a matter of 
debate, with the consequence that
SFT models continue to differ significantly in their outputs.
This is an unsatisfactory situation, considering how useful accurate and spatially resolved representations of surface magnetic flux evolution 
would be for data assimilation-based cycle prediction schemes
(e.g., \citealt{Kitiashvili2008,Dikpati2014}, and references therein).
Moreover, the availability of realistic, detailed surface magnetic maps 
associated with distinct dynamo regimes is needed in reconstructing the 
heliospheric magnetic field in the distant past 
(see, e.g., \citealt{Riley2015}).
In order to build a SFT model that behaves, as much as possible, like 
the Sun ---one to be ultimately used as the key surface
component of a solar-like 
Babcock--Leighton dynamo model---
calibration against observations needs to be performed thoroughly.
Some quantitative studies have been conducted (see, e.g.,
\citealt{Yeates2014}), but never through systematic
optimization procedures. This is what we aim to achieve in 
the present study.

We first discuss the formulation of the SFT model 
itself (\S~2), after which we turn to its calibration against observed 
data. Toward this end we used a genetic algorithm, which allows an 
efficient exploration of the model's parameter space, as well as the 
identification of parameter correlations and degeneracies (\S~3).
We then repeat the analysis while allowing the meridional flow to 
vary systematically in the course of the cycle, as suggested
by observations.
In \S~4 we explore the model behavior with respect to the 
stochastic variability inherent to emergence statistics.
We conclude by comparing and contrasting our optimized {\SFT} model
to similar models available in the extant literature.
Coupling to the dynamo simulation, and the resulting solar cycle model, 
is the subject of the following paper in this series 
(A. Lemerle \& P. Charbonneau 2015, in preparation).

\section{Model} \label{s_model}

As new {\BMR}s emerge at the surface of the Sun and subsequently decay,
their magnetic flux is dispersed and transported with the plasma by 
surface flows, and locally destroyed or amplified according to basic 
rules of electromagnetic induction.
For physical conditions representative of the solar photosphere, this 
process is well described by the MHD induction equation:
\begin{equation}
   \pd{\vec{B}}{t} = \curl (\vec{u} \cprod \vec{B} - \eta \curl \vec{B}) 
   \ , \label{eq_mhd}
\end{equation}
with $\eta$ the net magnetic diffusivity, including contributions from 
the small microscopic magnetic diffusivity $\eta_e = c^2/4\pi\sigma_e$
(with $\sigma_e^{-1}$ the electrical resistivity of the plasma), as well 
as a dominant turbulent contribution associated with the destructive 
folding of magnetic field lines by small-scale convective fluid motions.
A dynamically consistent approach would require Equation~(\ref{eq_mhd})
to be augmented by the hydrodynamical fluid equations including
Lorentz force and Ohmic heating terms.
However, on spatial scales commensurate with the solar radius, the use 
of a kinematic approximation, whereby the flow $\vec{u}$ is considered 
given, has been shown to be quite appropriate in reproducing the 
synoptic evolution of the solar surface magnetic field 
\citep[see, e.g.,][]{Wang2002a,Baumann2004}.
We adopt this kinematic approach in what follows, and solve 
Equation~(\ref{eq_mhd}) on a spherical shell representing the solar 
photosphere.

On spatial scales much larger than convection, only 
meridional circulation $\vec u_\text{P}(r,\theta)$ and
differential rotation $r\sin\theta\Omega(r,\theta) \uvec{e}_\phi$ 
contribute to $\vec{u}$ in Equation~(\ref{eq_mhd}).
Both these flows can be considered
axisymmetric ($\partial/\partial\phi\equiv 0$)
and steady ($\partial/\partial t\equiv 0$) 
to a good first approximation.
Since we solve the induction equation on the solar surface, 
meridional circulation reduces to a latitudinal flow
$\vec u_\text{P}\equiv u_\theta(R,\theta) \uvec{e}_\theta$.



Following earlier modeling work on surface magnetic flux evolution,
we consider the magnetic field to be predominantly radial on global 
scales and we solve only the $r$-component of Equation~(\ref{eq_mhd}),
after enforcing the null divergence condition throughout:
%
\begin{align}
   \pd{B_R}{t} =
   &- \frac{1}{R \sin\theta} 
   \pd{}{\theta}\big[ \sin\theta \, u_\theta(R,\theta) B_R \big] - 
   \Omega(R,\theta) \pd{B_R}{\phi} \nonumber \\
   &+ \frac{\eta_R}{R^2} \left[ \frac{1}{\sin\theta} 
   \pd{}{\theta}\left(\sin\theta\pd{B_R}{\theta}\right) + 
   \frac{1}{\sin^2\theta} \pdd{B_R}{\phi} \right] \nonumber \\
   &- \frac{B_R}{\tau_R} + S_{\BMR}(\theta,\phi,t) \ ,
   \label{eq_surftrans}
\end{align}
where $\eta_R$ is the uniform surface diffusivity.
Note the addition of two supplementary terms: a source term 
$S_{\BMR}(\theta,\phi,t) = \sum_i B_i(\theta,\phi)\delta(t-t_i)$, with 
$\delta$ the Dirac delta, to account for the emergence of new {\BMR}s at 
given positions $(\theta_i,\phi_i)$ and times $t_i$, 
to be extracted from some suitable observational database 
(see \S~\ref{s_W21}), and a linear sink term $-B_R/\tau_R$ to allow for 
some exponential decay of the surface field with time.
This term thus mimics the radial diffusion and mechanical subduction
of locally inclined magnetic field lines, which cannot be captured 
by Equation~(\ref{eq_surftrans}) and the assumption of a purely radial 
surface magnetic field.
The addition of this sink term is also motivated by the analysis of 
\citet{Schrijver2002}, who found that such decay on a timescale of 
$\unit{5-10}{\years}$ was necessary to preclude secular drift and ensure 
polarity reversal of the polar caps when modeling surface flux evolution 
over many successive cycles.
\citet{Baumann2006} argued that this exponential destruction of surface
magnetic flux could be justified physically as the effect of a vertical
turbulent diffusion (including convective submergence) on the decay of
the dominant dipole mode.
In what follows we treat $\tau_R$ as a free parameter.
Equation~(\ref{eq_surftrans}) is now a two-dimensional linear
advection--diffusion equation for the scalar component 
$B_R = B_r(R,\theta,\phi,t)$ at the surface of the Sun, augmented by 
source and sink terms.

\subsection{Meridional circulation} \label{s_mercirc}

Because the solar meridional surface flow is weak and thus
difficult to measure accurately
(but do see \citealt{Ulrich2010}), its latitudinal dependence has been
approximated by a number of ad hoc analytical formulae:
some as minimalistic as a $\cos\theta\sin\theta$, with peak at 
$\unit{45}{\degree}$ latitude \citep[e.g.][]{Dikpati1999},
some displacing the peak flow to lower latitudes by introducing
exponents to the $\cos\theta$ and $\sin\theta$ terms
\citep[e.g.][]{vanBalle1988,Wang2002b},
others using a truncated series expansion \citep{Schrijver2001},
or shutting down the flow speed to zero near the poles
\citep[e.g.][]{vanBalle1998}, for a closer fit to the observed motion
of surface magnetic features 
\citep{Komm1993,Hathaway1996,Snodgrass1996}. 

\begin{figure}
   \plotone{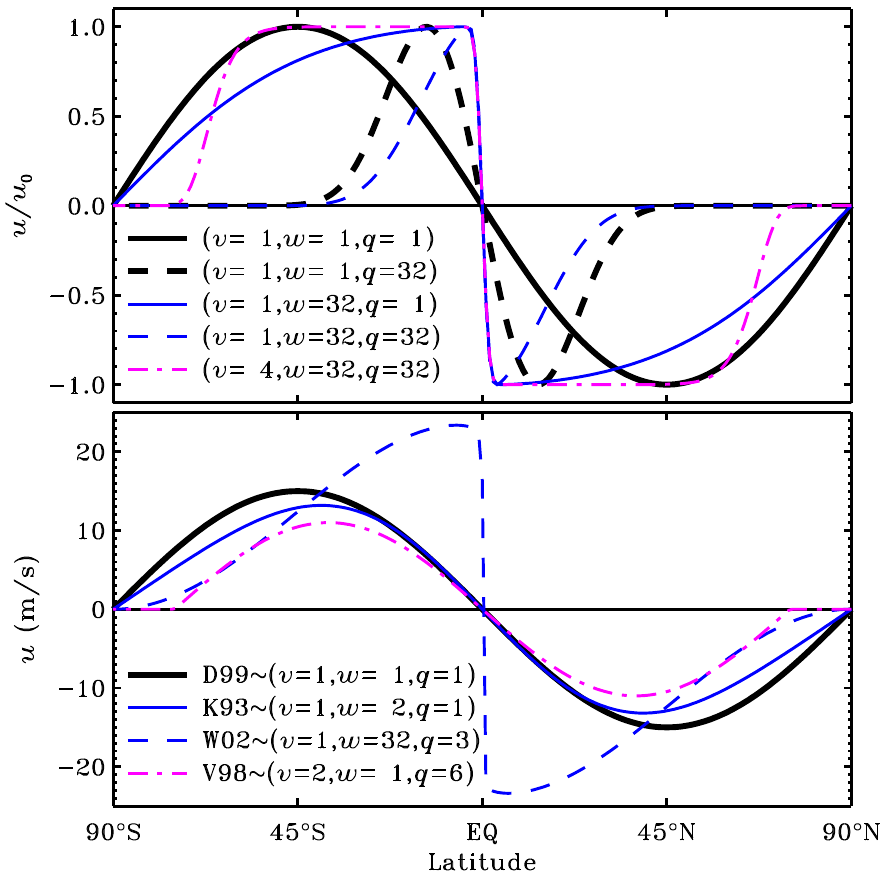}
   \caption{Top panel: sample profiles of surface meridional 
   circulation, 
   as formulated in Equation~(\ref{eq_mercirc}). Bottom panel: other 
   profiles found in the literature: D99 \citep{Dikpati1999}, 
   K93 \citep{Komm1993}, W02 \citep{Wang2002b}, and 
   V98 \citep{vanBalle1998}, with corresponding parameter values 
   when approximated using Equation~(\ref{eq_mercirc}).}
   \label{f_mercirc}
\end{figure}

The recent observational determinations of \citet{Ulrich2010}, however, 
suggest the existence of a more complex latitudinal pattern, 
increasing quite rapidly from the equator to a peak amplitude 
near $\unit{15}{\degree}$ to $\unit{20}{\degree}$ latitude,
and decreasing more slowly to zero up to 
$\unit{60}{\degree}$--$\unit{70}{\degree}$ latitude. 
To account for such asymmetric rise and fall of the flow speed at 
low--mid latitudes and possible suppression of the flow at high 
latitudes, we opt to use the following, versatile analytical formula:
\begin{equation}
   u_\theta(R,\theta) = -u_0 ~ 
   \erf^q\bigl(v\sin\theta\bigr) ~\erf^n\bigl(w\cos\theta\bigr) \ ,
   \label{eq_mercirc}
\end{equation}
with $u_0$ the maximum flow velocity and $q$, $n$, $v$, and $w$ free 
parameters to be determined in the course of the foregoing analysis.
With $u_0$ the same in both hemispheres, the profile is antisymmetric 
with respect to the equator.
It takes approximately the shape of a $\cos\theta\sin\theta$ profile in 
the case $q=1$, $n=1$, $v=1$, and $w=1$, with peak at mid--latitudes.
Varying parameters $w$ and $v$ allows the latitude of peak flow speed to 
be moved to either lower latitudes ($w>1$) or higher latitudes ($v>1$). 
High values for both $w$ and $v$ broaden the peak between low and high 
latitudes. 
Values of $q>1$ have the effect of stopping the flow before the poles, 
at lower latitudes as $q$ increases.
Growing values of $n$ have the same effect near the equator, but since 
such a low-latitude $\unit{0}{\mps}$ plateau seems far from a solar 
behavior, we set $n=1$ for the remainder of our analysis.
The top panel of Figure~\ref{f_mercirc} illustrates a few sample
profiles. Note in particular that, 
with appropriate choices for $q$, $v$ and $w$, Equation~(\ref{eq_mercirc})
can reproduce most profiles in use in the literature (see bottom panel).

Analyses by \citet{Ulrich2010} also show latitudinal flow speeds 
dropping to negative values near the poles, especially at the beginning 
and end of cycles.
Since this pattern is at the limit of observational determinations and
does not appear in all cycles or solar hemispheres, we opt not to 
model these potential high-latitude secondary flow cells.
We also use the same value of $u_0$ in both hemispheres.

\subsection{Differential rotation} \label{s_diffrot}

Unlike meridional circulation, the surface differential rotation profile
is observationally well established. We adopt here the parametric
formulae calibrated helioseismically by \citet{Charbonneau1999}:
\begin{equation}
   \Omega(R,\theta) = 
   \Omega_0 \left( 1 + a_2 \cos^2\theta + a_4 \cos^4\theta \right) \ ,
   \label{eq_surfdiffrot}
\end{equation}
with $a_2=-0.1264$ and $a_4=-0.1591$.
The angular velocity is lowest at the poles and highest at the equator,
where $\Omega_0 = \unit{2.894}{\micro\radps}$ 
(see also \citealt{Snodgrass1983}).

\subsection{Magnetic diffusivity} \label{s_magdiff}

Near the solar surface, the magnetic diffusivity due to Ohmic 
dissipation reaches $\eta_e\simeq\unit{10^7}{\cmmps}$.
However, as initially shown by \citet{Leighton1964}, surface convective 
motions at the supergranular scale drive a random walk that disperses 
magnetic flux, and can be modeled as a diffusive process characterized 
by an effective magnetic diffusivity of order 
$\eta_R\simeq\unit{10^{12}-10^{13}}{\cmmps}$.
The exact value is virtually impossible to determine from first 
principles, so we henceforth treat $\eta_R$ as a free parameter to be 
determined by our analysis.

\subsection{Numerical solution} \label{s_grid}

The adimensional form of the surface transport 
Equation~(\ref{eq_surftrans}) is solved numerically in a two-dimensional 
Galerkin finite-element scheme (see, e.g., \citealt{Burnett1987}), over 
a regular Cartesian grid in $[\theta,\phi]$ with longitudinal 
periodicity, the latter enforced through a padding of ghost cells, 
updated at every time step.
The zero-flux polar boundary condition ${dB_R}/{d\theta} = 0$ is
hardwired at the level of the finite-element scheme itself.
Because only a small fraction of emerging magnetic flux ends up
accumulating at the poles, it is essential to rigorously ensure
magnetic flux conservation.
Consequently, the numerical discretization errors must be monitored and 
kept in check.

Using double precision arithmetic, a 
$N_\phi{\times}N_\theta = 256{\times}128$ longitude--latitude grid is 
required to ensure that net signed surface flux never exceeds $10^{-2}$ 
of the total unsigned surface flux, at least an order of magnitude
better than observations (see, e.g., Figure~\ref{f_cyc21}e).
With such a grid, relatively short time steps are also necessary to
ensure stability, which is of the order of $100$ time steps per year, 
for a total of $N_t\simeq10^3$ time steps per solar cycle.

\subsection{Numerical optimization} \label{s_opt}

The final formulation of Equation~(\ref{eq_surftrans}) leaves us with a 
set of six unknown parameters 
($u_0$, $q$, $v$, $w$, $\eta_R$, and $\tau_R$),
which we aim to constrain quantitatively based on magnetographic 
observations of the solar surface.

We opt to compare a time--latitude map of longitudinally averaged
$\langle B_R \rangle^\phi (\theta,t)$ output from our model with an 
equivalent longitudinally averaged magnetogram.
For this purpose, \citet{Hathaway2010} has provided us with his 
well-known ``magnetic butterfly diagram'' data\footnote{
\url{http://solarscience.msfc.nasa.gov/images/magbfly.jpg}},
made available from 1976 August to 2012 August, at a temporal 
resolution of one point per Carrington rotation and $180$ data points 
equidistant in $\cos\theta$.
The data are a compilation of measurements from instruments on Kitt Peak 
and SOHO,
corrected to obtain the radial component of the magnetic field.
%

\begin{figure}
   \plotone{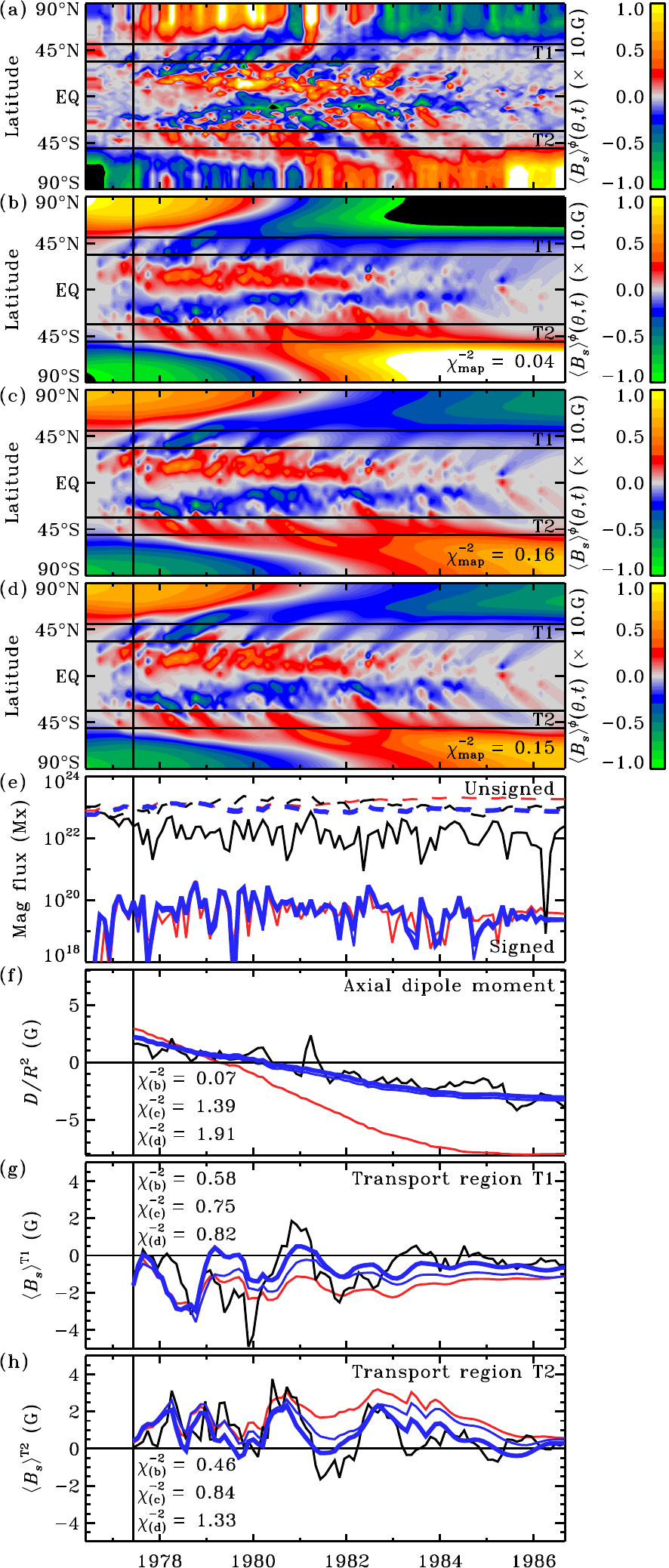}
   \caption{Time--latitude contour plots of the radial surface magnetic 
   field from 
   (a) magnetographic data by \citet{Hathaway2010}, cycle 21 only,
   (b) an unacceptable solution using the reference values listed in 
   Table~\ref{t_param} ($\chi^{-2}=0.09$), 
   (c) an example of a suboptimal but acceptable solution with 
   $\chi^{-2}=0.42=93\%\chi^{-2}_\text{max}$, and 
   (d) an optimal solution ($\chi^{-2}_\text{max}=0.45$).
   Temporal evolution of 
   (e) total unsigned (dashed) and absolute signed (continuous) magnetic 
   flux, 
   (f) axial dipole moment, and average radial magnetic field in 
   (g) transport region T1 (latitudes $\unit{34}{\degree}$N to 
   $\unit{51}{\degree}$N), and 
   (h) transport region T2 (latitudes $\unit{34}{\degree}$S to 
   $\unit{51}{\degree}$S).
   Thin black curves are extracted from the data presented in (a), 
   thin red from the unacceptable solution (b), 
   thin blue from the suboptimal solution (c), 
   and thick blue from the optimal solution (d). 
   The vertical straight line delineates the first $10$\% of the 
   simulation time, which is excluded from the fitness calculation.
   Also listed are the values of sub-criteria $\chi^{-2}_\text{map}$, 
   $\chi^{-2}_D$, $\chi^{-2}_\text{T1}$, and $\chi^{-2}_\text{T2}$ for 
   each solution.}
   \label{f_cyc21}
\end{figure}

As a unique optimization criterion for the model, a plain minimization 
of the residuals between the two maps represents the most 
straightforward approach.
Unfortunately, this turns out to be insufficient to properly constrain
the model parameters, for a number of reasons including high-latitude
artifacts, as well as observational size and magnetic field thresholds
leading to missing flux.
%
Therefore, in addition to fitting the time--latitude synoptic map,
we add further constraints by putting more weight on two 
physically meaningful features:
first, the evolution of the overall axial dipole moment 
(divided by $R^2$), defined as
\begin{equation}
   D^\ast(t) =\frac{D(t)}{R^2} = \frac{3}{2} \int_0^\pi 
   \langle B_R \rangle^\phi (\theta,t) \cos\theta \sin\theta d\theta \ ;
\end{equation}
%
and, second, the shape of mid--latitudes flux-migration strips.
To quantify the latter, we delimit two ``transport regions'',
one in each hemisphere (regions T1 and T2 in Figure~\ref{f_cyc21}a,
latitudes $\pm\unit{34}{\degree}$ to $\pm\unit{51}{\degree}$), dominated 
by inclined flux strips and where very little flux emerges or 
accumulates.
We then calculate the latitudinal average of $B_R$ in each region, a 
quantity directly influenced by the amount of flux and the width and 
inclination of flux strips:
\begin{equation}
   \langle B_R \rangle^\text{T1,T2}(t) = \frac{ \int_\text{T1,T2} 
   \langle B_R \rangle^\phi (\theta,t) \sin\theta d\theta}
   { \int_\text{T1,T2} \sin\theta d\theta} \ .
\end{equation}
%
With such definitions, both of the above surface integrals end up having 
the same physical unit, with magnitude of the order of a few Gauss.

Constraining the behavior of the axial dipole component, and indirectly 
the times of polarity reversals,
should help constrain the values  of 
diffusivity $\eta_R$ and exponential decay time $\tau_R$, which directly 
shape the polar magnetic caps,
as well as meridional circulation parameters ($u_0$, $q$, $v$, and $w$), 
which dictate how new flux migrates toward the poles and
eventually triggers the polarity reversals. The extent of polar 
caps down to $\unit{60-70}{\degree}$ latitude also suggests a 
significant decrease of the meridional flow near these high latitudes.
Similarly, diffusivity will shape the width and length of mid--latitudes 
flux-migration strips,
and the meridional circulation profile will set their inclination
in the synoptic map.

To obtain a final optimization criterion, we evaluate 
the rms deviation $\chi_\text{map}$ between
simulated $\langle B_R \rangle^\phi_\text{sim} (\theta,t)$ map and 
measured $\langle B_R \rangle^\phi_\text{dat} (\theta,t)$ map, 
the rms deviation $\chi_D$ between 
simulated $D^\ast_\text{sim}(t)$ and 
measured $D^\ast_\text{dat}(t)$, and 
the rms deviations $\chi_\text{T1,T2}$ between 
simulated $\langle B_R \rangle^\text{T1,T2}_\text{sim}(t)$ and 
measured $\langle B_R \rangle^\text{T1,T2}_\text{dat}(t)$.
We combine them as follows, such that 
the overall rms deviation $\chi$ must be minimized:
\begin{equation}
   \chi^2 = \frac{1}{4} \Bigr(\chi^2_\text{map} + \chi^2_D + 
   \chi^2_\text{T1} + \chi^2_\text{T2} \Bigl) \ .
   \label{eq_fitns}
\end{equation}
%
This multi-objective optimization criterion goes significantly beyond 
the cross-correlation approach introduced by \citet{Yeates2014}, since 
it defines an absolute least-squares minimization of the differences 
between the model and observations, rather than being restricted to 
their temporal synchronisation.

The inverse of the quantity $\chi^2$ is defined as our merit function, 
or ``fitness''.
We seek to maximize this fitness using the genetic algorithm-based
optimizer PIKAIA 1.2,
a public domain software distributed by HAO/NCAR\footnote{
\url{http://www.hao.ucar.edu/modeling/pikaia/pikaia.php} (2015 March)}.
Genetic algorithms ({\GA}) are a biologically inspired class of 
evolutionary algorithms that can be used to carry out global numerical 
optimization.
PIKAIA \citep{Charbonneau1995,Charbonneau2002b} is one such classical 
{\GA}-based optimizer.
PIKAIA evolves an optimal solution to a given optimization task by 
selecting the better solutions among a population of trial solutions, 
and breeding new solutions through genetically inspired operations of 
crossover and mutation acting on a string encoding of the selected 
solution's defining parameters.
In this manner {\GA} allow efficient, adaptive exploration of parameter 
space through parallel processing of advantageous substrings. 
Indeed, GA-based optimizers have proven quite robust in handling global
optimization problems characterized by complex, multimodal parameter
spaces that often trap gradient-based optimizers in local extrema.
For an accessible introduction to GA and their 
use for numerical optimization, see \citet{Charbonneau2002a}.

In the present context PIKAIA is operating in a seven-dimensional
parameter space (see Table~\ref{t_param}), with the fitness measure 
given by Equation~(\ref{eq_fitns}).
We use the default settings for PIKAIA's internal control parameters, 
with the exception of encoding depth, population size, mutation 
mode (equiprobable digit$+$creep, with fitness-based adjustment),
and number of generational iterations. 

As numerical optimization algorithms, {\GA} tend to be computationally
expensive, as the number of model evaluations is equal to the population
size times the number of generational iterations.
In most model fitting tasks reported upon in what follows, a population
size of 48 trial solutions evolving over 500 generations was found to
be sufficient to reliably ensure
proper convergence of all model parameters. This adds up to 24000 model
fitness evaluations per optimization run.
Calculating the fitness of a single trial solution (seven-parameter 
vector) implies running a {\SFT} simulation, calculating
the various integrals introduced in \S~\ref{s_opt}, and finally 
evaluating Equation~(\ref{eq_fitns}).
For our working spatial mesh and time step this requires a little under 
10 minutes on a single-core modern CPU, adding
up to $167$ core-days for a typical optimization run.
However, this sequence of operations is applied independently to each 
member of the population, and so can easily be carried out in parallel 
(see, e.g., \citealt{Metcalfe2003}).
With the only information returned by each evaluation being the fitness, 
near-perfect parallelization can be achieved, by assigning one core per 
population member, thus bringing the wall-clock time down to a few days.

One specific feature of PIKAIA deserves further discussion, namely
its adaptive mutation rate. Throughout the evolution, PIKAIA monitors
the fitness differential and spread of the population in parameter 
space, and whenever these quantities become too small (large), the 
internal mutation rate is increased (decreased), while ensuring that
the current best individual is always copied intact into the next
generation.
This effectively leads to a form of Monte Carlo exploration of parameter 
space in the vicinity of the current optimum, allowing escape from local 
extrema as the case may be.
At the same time, this pseudo-random sampling of parameter space taking 
place about the current optimum solution can be harnessed to construct
error estimates on solution parameters; 
these estimate remain useful even though, strictly speaking, population 
members are not statistically independent of one another since they 
have all been bred from the same earlier generations.
%

\section{Solar cycle 21: a case study} \label{s_cyc21}

\subsection{Surface emergence database} \label{s_W21}

In an effort to characterize the details of surface flux evolution 
during sunspot cycle 21, \citet{Wang1989-0} (hereafter {\WS}) have 
assembled a comprehensive database of over 3000 observed {\BMR}s, each 
approximated as a pair of poles of identical magnetic flux but opposite 
polarity.
Their input data consisted of daily magnetograms recorded at the 
National Solar Observatory/Kitt Peak between 1976 August and 1986 April. 
For each {\BMR}, they list the time, magnetic flux, polarity of the 
western pole, and latitude and longitude of each pole, measured
when their magnetic flux reached its peak.
Positive magnetic fluxes range from \unit{10^{20}}{\maxwell} to 
\unit{7\times10^{22}}{\maxwell}, for a total of
$\unit{1.09\times10^{25}}{\maxwell}$ for the whole database.
As described in {\WS}'s analysis, while the 
latitude and time of emergence of the {\BMR}s follow the usual butterfly 
pattern, the polarities of western poles are predominantly coherent in a 
given hemisphere and opposite in the other hemisphere, as per Hale's 
law.
The database also shows a slight hemispheric asymmetry of $0.4\%$ in 
favor of the northern hemisphere, in both total flux and number of 
{\BMR}s.

We use {\WS}'s updated database entries as direct inputs for the 
{\SFT} source term $S_\BMR(\theta,\phi,t)$.
Each $i$th {\BMR} is injected in the surface layer at its observed time 
$t_i$, colatitude $\theta_i$, and longitude $\phi_i$, with a gaussian 
distribution for each pole:
\begin{equation}
   B_i(\theta,\phi) = 
   \underbrace{ B_{i0}e^{-\delta_{i+}^2/2\sigma^2}}
   _{B_{i+}(\theta,\phi)} 
   +
   \underbrace{-B_{i0}e^{-\delta_{i-}^2/2\sigma^2}}
   _{B_{i-}(\theta,\phi)} 
   \ ,
\end{equation}
where $\delta_{i+}$ and $\delta_{i-}$ are the heliocentric angular 
distances from  the centres $(\theta_{i+},\phi_{i+})$ and 
$(\theta_{i-},\phi_{i-})$ of the two poles, respectively, and 
$\sigma=\unit{4}{\degree}$ the width of the gaussian. 
We set this width to a fixed value for all emergences, in order to 
minimize numerical errors, a choice that may induce only slight shifts 
in the times of emergences. In other words, our Gaussian-shape BMRs are 
injected with a larger area than observed, but this does not impact the 
subsequent evolution since Gaussian profiles spread in a 
shape-preserving manner under the sole action of diffusion.
Also, the use of heliocentric distances ensures that the surface 
integral of $B_{i+}(\theta,\phi)$ can be calculated exactly as the 
measured flux, provided that magnetic field amplitude $B_{i0}$ at the 
centre of the gaussian is adjusted accordingly.
%
%
This appropriate geometry also guarantees that total signed flux from 
both poles cancels completely.
This condition is important since any remnant net surface flux was found 
to accumulate during the simulation.
All these factors considered, along with a tight grid and 
double precision arithmetic, the net magnetic flux from one discretized 
{\BMR} rarely exceeds $10^{-13}$ times the emerged flux in the positive 
pole.

\subsection{Initial condition}

\begin{figure}
   \plotone{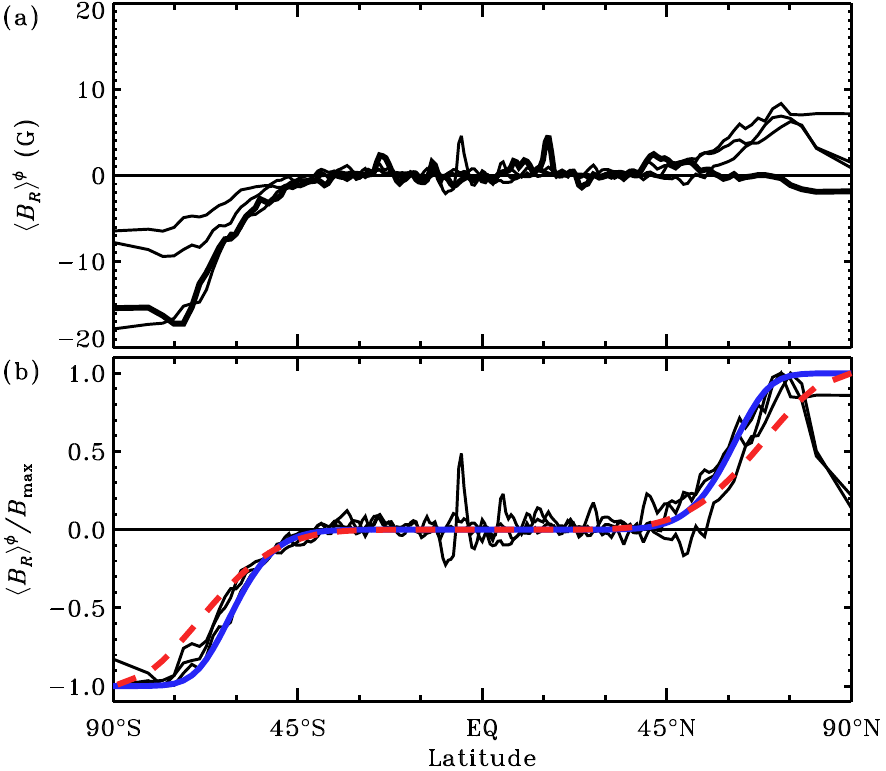}
   \caption{(a) Latitudinal distributions of $B_R$ during the last four 
   activity minima: 1976 June (thick black) and 1986 September, 1996 
   May, and 2008 December (thin black).
   (b) Same curves, for the last three minima, but normalized to $1$ in 
   each hemisphere.
   Superimposed are profiles~\ref{eq_initcnd2} (thick blue) and 
   \ref{eq_initcnd1} (thick red dashed).}
   \label{f_initcnd}
\end{figure}

As we aim to reproduce, among other global cycle features, the timing of 
polar cap polarity reversals, the amount and distribution of magnetic 
flux at the beginning of the cycle is particularly crucial. 
Even if we assumed an axisymmetric initial condition, the synoptic data 
compiled by \citet{Hathaway2010} remain incomplete for this 1976-1977 
activity minimum.
In fact, while the southern hemisphere presents the expected quantity of 
negative flux, the northern hemisphere presents no clear polar cap, and 
even some substantial negative flux through the end of 1976 
(see Figure~\ref{f_cyc21}a).
Figure~\ref{f_cyc21}e illustrates how the corresponding signed magnetic 
flux is far from zero for this period, even reaching the value of 
unsigned flux at one point. Figure~\ref{f_initcnd}a illustrates the 
latitudinal distribution of $B_R$ in 1976 June, patently not 
mirror-symmetric about the equatorial plane.

In order to construct a plausible initial condition, we turn to observed 
distributions for other cycle minima.
Figure~\ref{f_initcnd}b illustrates latitudinal distributions of $B_R$ 
at the beginnings of cycles 22--24, normalized to unity in each 
hemisphere. 
Apart from the unusually asymmetric profile for cycle 21,
the latitudinal distributions are characterized by a low field strength 
from the equator to $\simeq\unit{45}{\degree}$ latitude, followed by a 
rise to maximum field strength near $\unit{70}{\degree}$ latitude, in 
both hemispheres. 
The amplitude appears to fall near the poles for some activity minima, 
but we do not take this feature into account considering the lower 
reliability of high-latitude line-of-sight magnetographic measurements. 
While the simple axisymmetric profile
\begin{subequations}
\begin{equation}
   B_R(\theta,\phi,t_0) = B_0 \abs{\cos\theta}^7 \cos\theta
   \label{eq_initcnd1}
\end{equation}
has been adopted by many authors \citep[e.g.][]{Svalgaard1978,Wang1989}, 
a curve of the form
\begin{equation}
   B_R(\theta,\phi,t_0) = B_0^\ast 
   \erf\biggl(\frac{\abs{\cos\theta}^{11} \cos\theta}{\pi/8}\biggr) 
   \label{eq_initcnd2}
\end{equation}
\end{subequations}
fits better the measurements shown in Figure~\ref{f_initcnd}. 
We test both functional forms in what follows.
In both cases, the same amplitude $B_0$ is used in the two hemispheres, 
to ensure a zero net flux, but is kept as a seventh free parameter to be 
optimized along with the six physical parameters described above.

Finally, for the same reason that prevents us from using 1976 June 
measurements as an initial condition, we should probably not expect to 
reproduce the synoptic magnetogram for early cycle 21. We therefore 
choose to perform our optimization over the period going from 1977 June 
to 1986 Septembre.

\subsection{Reference case}

As a benchmark for subsequent comparison, we first look at the behavior 
of our model using some typical parameter values found in the 
literature. 
\citet{Baumann2004} presented a detailed study of the effect of varying 
parameters on a {\SFT} model similar to ours.
For their reference case, they used a diffusivity 
$\eta_R=\unit{600}{\kmmps}$ and a maximum meridional flow velocity 
$u_0=\unit{11}{\mps}$, with the profile by \citet{vanBalle1998}, as 
illustrated in Figure~\ref{f_mercirc}. 
This profile can be closely approximated by our adopted profile
(Equation~(\ref{eq_mercirc})), by setting $q=6$, $v=2$, and $w=1$. 
They did not use any exponential decay term, which is similar to 
adopting $\tau_R \gtrsim \unit{32}{\years}$ in Equation~(\ref{eq_surftrans}). 
Figure~\ref{f_cyc21}b illustrates the results of our {\SFT} 
simulation in such a parameter regime (listed in the first 
column of Table~\ref{t_param}), starting with the
$\unit{11.5}{\gauss} \abs{\cos\theta}^7 \cos\theta$ initial condition by 
\citet{Svalgaard1978}.

Visual comparison of panels (a) and (b) of Figure~\ref{f_cyc21} is 
encouraging, with some typical mid--latitude flux-migration strips of 
the right polarity, and accumulation of polar flux at about the right 
time and amplitude. 
However, the rms deviation between this 
time--latitude map and its observed counterpart (Figure~\ref{f_cyc21}a) 
is $\chi_\text{map}= \unit{4.9}{\gauss}$, which is quite high. 
Moreover, an excessive 
quantity of opposite magnetic flux accumulates in polar regions during 
the second half of the cycle, with final unsigned magnetic flux reaching 
about twice the observed value of $10^{23}{\maxwell}$ at the end of the 
cycle (cf.~black and red dashed curves in Figure~\ref{f_cyc21}e). 
The comparison worsens further when looking at the evolution of the 
axial dipole moment (Figure~\ref{f_cyc21}f): the simulated curve diverges 
markedly from observations, with $\chi_D= \unit{3.9}{\gauss}$. 
Again, the initial axial dipole is too strong, but this does not prevent 
an early reversal and the building of dipolar moment in excess of the 
observed value by over a factor of two.
This excessive transport of flux toward the poles is presumably due to 
a combination of a slow poleward flow at low latitudes, allowing too 
much of the polarity of the {\BMR}s' western poles to cancel at the 
equator, and a fast latitudinal transport of the remaining polarity at 
mid/high latitudes. 
This suggests the use of a suboptimal meridional circulation profile.

The detailed shape of flux-migration strips, evaluated using the 
integrated magnetic field in the transport regions T1 
(Figure~\ref{f_cyc21}g) and T2 (Figure~\ref{f_cyc21}h), does not compare 
well to observations either ($\chi_\text{T1}= \unit{1.3}{\gauss}$ and 
$\chi_\text{T2}= \unit{1.5}{\gauss}$).
The subdued variability reveals a temporal widening of the flux strips, 
likely due to an excessive surface magnetic diffusivity. 
Altogether, these features lead to a combined rms difference 
$\chi=\unit{3.3}{\gauss}$ between this reference case and observations 
($\chi^{-2}=0.09$).

\begin{deluxetable}{lrcl}
\tablecaption{Optimal Parameter Values\label{t_param}}
\tablehead{
\colhead{Symbol}& \colhead{Reference}   & \colhead{Tested}          & \colhead{Optimal}                                                                               \\*[0.0ex]
                & \colhead{Value}       & \colhead{Interval}        & \colhead{Solution}                                                                              \\*[0.0ex]
          & \colhead{($\chi^{-2}=0.09$)}& \colhead{}                & \colhead{($\chi^{-2}\in[0.42,0.45]$)}                                                                    }
\startdata
\quad$B_0$      &          $11.5\qquad$ & $[0\phm{0^0},25\phm{^0}]$ & $\tblmk{\phm{b}\phm{c}}\phm{0}8.5\pm\sstk{3.5\\2.5}\phm{00}$G                 \\*[1.0ex]
\quad$\tau_R$   & \tblmk{a}$~~32\qquad$ & $[2^1\phm{0},2^5\phm{0}]$ & $\tblmk{\phm{b}\phm{c}}\phm{.0}32\pm\sstk{\8\phm{.}\\25\phm{.}}\phm{00}$\years\\*[1.0ex]
\quad$w$        &             $1\qquad$ & $[1\phm{0^0},2^5\phm{0}]$ & $\tblmk{\phm{b}\phm{c}}\phm{0.0}8\pm\sstk{24\phm{.}\\4\phm{.0}}\phm{00}$      \\*[1.0ex]
\quad$v$        &             $2\qquad$ & $[1\phm{0^0},2^3\phm{0}]$ & $\tblmk{\phm{b}\phm{c}}\phm{0}2.0\pm\sstk{1.5\\1.0}\phm{00}$                  \\*[1.0ex]
\quad$q$        &             $6\qquad$ & $[1\phm{0^0},2^5\phm{0}]$ & $\tblmk{\phm{c}\phm{}b}\phm{0.0}7\pm\sstk{4\phm{.0}\\3\phm{.0}}\phm{00}$      \\*[1.0ex]
\quad$u_0$      &            $11\qquad$ & $[5\phm{0^0},30\phm{^0}]$ & $\tblmk{\phm{}\phm{b}c}\phm{0.}12\pm\sstk{4\phm{.0}\\2\phm{.0}}\phm{00}$\mps  \\*[1.0ex]
\quad$\eta_R$   &           $600\qquad$ & $[10^2\phm{},10^4\phm{}]$ & $\tblmk{\phm{c}\phm{}d}\phm{.}350\pm70\phm{\sstk{0.0\\0.0}}$\kmmps
\enddata
\tablenotetext{}{\textbf{Notes.}}
\tablenotetext{a}{$\tau_R \gtrsim \unit{32}{\years}$ is similar to 
removing the term $-B_R/\tau_R$ in Equation~(\ref{eq_surftrans}).}
\tablenotetext{b}{Solution for $q$ when $v{=}2$. Otherwise 
$q{=}{\left(2.8\pm\substack{2.0\\1.1}\right)}\cdot{2^{1.25(\log_2v)^2}}$ 
(optimization W21-7; see Figure~\ref{f_fitlandqv}).}
\tablenotetext{c}{Solution for $u_0$ when $v{=}2$, $q{=}7$, and $w{=}8$
(optimization W21-2). 
Overall solution: $u_0\in\unit{[8,18]}{\mps}$ (optimization W21-7).}
\tablenotetext{d}{Solution for $\eta_R$ when $u_0{=}\unit{12}{\mps}$, 
$v{=}2$, $q{=}7$, and $w{=}8$.
More generally, 
$\eta_R=\unit{(350\pm70)\cdot10^{0.037(u_0-12)}}{\kmmps}$ when $v{=}2$, 
$q{=}7$, and $w{=}8$ (optimization W21-2; see Figure~\ref{f_fitland2d}).
Overall solution: $\eta_R\in\unit{[240,660]}{\kmmps}$ 
(optimization W21-7).}
\end{deluxetable}

\subsection{Optimal solution} \label{s_optsol}

We perform our main optimization of the cycle 21 simulation and its 
seven-parameters set (hereafter W21-7), based on fitness $\chi^{-2}$ 
(Equation~(\ref{eq_fitns})).
For each parameter, we choose an interval to be explored that is both 
physically meaningful and numerically stable.
For instance, values of $q>32$ and $w>32$ tend to generate numerical 
instabilities due to excessive latitudinal gradients.
Surface diffusivity $\eta_R<\unit{10^2}{\kmmps}$ sometimes causes
problems when used in conjunction with our axisymmetric transport 
dynamo model of the solar interior.
Maximum flow speed $u_0\in\unit{[5,30]}{\mps}$ broadly corresponds to 
observations \citep[e.g.][]{Ulrich2010}, and similarly for 
$B_0\in\unit{[0,25]}{\gauss}$ (see Figure~\ref{f_initcnd}a).
Finally, the linear term $-B_R/\tau_R$ in Equation~(\ref{eq_surftrans}) has 
virtually no effect for $\tau_R\gtrsim\unit{32}{\years}$.
The intervals explored for each parameter are listed in the 
second column of Table~\ref{t_param}.

Figure~\ref{f_cyc21}d illustrates one optimal solution, with maximum 
fitness $\chi^{-2}_\text{max}=0.45$ ($\chi=\unit{1.5}{\gauss}$), 
significantly better than the $\chi^{-2}=0.09$ 
($\chi=\unit{3.3}{\gauss}$) obtained for the reference case. 
The general shape of the time--latitude map is visually similar to the 
reference case, but with a later polar field reversal, a magnetic cap 
slightly more confined to high latitudes, and a more reasonable maximum 
polar field of order $\unit{10}{\gauss}$ at the end of the cycle. 
This translates to a rms residual 
$\chi_\text{map}= \unit{2.6}{\gauss}$, a reduction by a factor of nearly
two as compared to the reference case of Figure~\ref{f_cyc21}b. 
The overall unsigned magnetic flux at the end of the cycle, where 
observed signed flux reaches its lowest value, is also closer to the 
observed $10^{23}{\maxwell}$ (Figure~\ref{f_cyc21}e). 
These improvements are even more obvious in terms of the axial dipole 
moment (Figure~\ref{f_cyc21}f), with a curve that nicely fits the general 
trend of observations ($\chi_D= \unit{0.72}{\gauss}$). 
Mid--latitudes flux-migration strips are better defined and distinct 
from one another, with appropriate time--latitude inclinations. 
This translates into more pronounced oscillations of the integrated 
magnetic field in transport regions T1 (Figure~\ref{f_cyc21}g) and T2 
(Figure~\ref{f_cyc21}h), in much better agreement with observed curves 
($\chi_\text{T1}= \unit{1.1}{\gauss}$ and 
$\chi_\text{T2}= \unit{0.87}{\gauss}$).

Nonetheless, despite this formal quantitative optimization, 
our best solution is unable to reproduce many details of the 
observed magnetic butterfly diagram. 
This means that high-frequency variations of polar magnetic fields must 
be either artefacts of high-latitude observations or, if real, would 
require the sporadic injection of magnetic flux opposite to the main 
trend, potentially from small flux strips, which are obviously not 
present in the simulations. 
On the other hand, most large flux strips are reproduced by the 
simulation, but not necessarily at the right moments. 
Looking in detail at transport region T1, we see:
a large negative strip that crosses the region during year 1978, 
well reproduced by the simulation; 
a second strip building progressively from early 1979 up to a peak at 
the end of the same year, triggering a first polar flux reversal in the 
northern hemisphere, but which fails to be reproduced with the correct 
amplitude; 
followed by a large positive strip in early 1981, which reverses back 
the polar cap field but is only slightly visible in the simulation; 
followed again, during year 1982, by a surge of negative flux that 
triggers the final polar field reversal, and finds its equivalent in 
two small strips in the simulation; 
and a small positive strip, which does not manage to reach 
mid--latitudes in the simulation.
Transport region T2 is somehow more satisfying with a series of five 
distinct positive strips from 1978 to 1985, reproduced almost exactly 
at the right moments, but interrupted by at least three negative strips 
only faintly visible in the simulation.
All these differences suggest that the simulation may not always 
transport the flux adequately from low latitudes.
This could be due to a missing time dependence in the meridional flow, 
due, e.g., to nonlinear magnetic feedback from activity bands 
\citep{Cameron2014},
or to oversimplifications in our emergence procedure,
in particular the slight temporal shifts induced by the use of fixed 
angular sizes for all emergences or the lower limit on BMR 
detection in {\WS}'s database.
The use of the SFT approximation itself also has obvious limits, in
particular the assumptions of a systematic radially oriented magnetic 
field and of uniform diffusion rate, the latter simplification breaking
down at the small scales where the advective motions of the 
magnetic elements occur, which can result in a small- to large-scale 
build-up of magnetic structures
(see, e.g., \citealt{Schrijver2001}).

\subsection{Parameter Analysis} \label{s_results21}

\begin{figure}
   \plotone{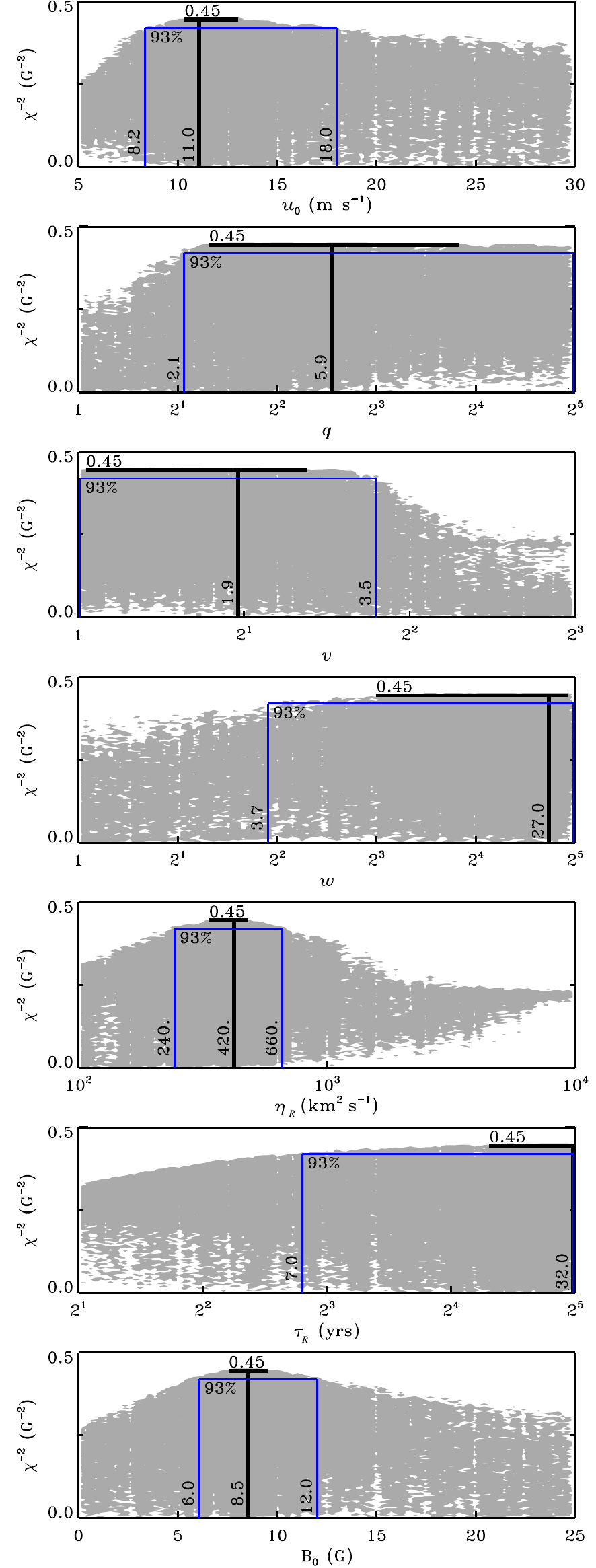}
   \caption{Fitness $\chi^{-2}$ as a function of parameter values, for 
   each of the $144000$ solutions obtained from optimization W21-7 
   (initial condition~\ref{eq_initcnd1}).
   On each plot, the thick horizontal line indicates the interval where 
   $\chi^{-2}\geq0.445$, and the thick vertical line the parameter value 
   where true maximum fitness $\chi^{-2}_\text{max}=0.450$ is reached.
   Thin vertical blue lines indicate the parameter values where fitness 
   reaches $93\%\chi^{-2}_\text{max}$, such that any solution above the 
   horizontal blue line is considered acceptable.}
   \label{f_fitland}
\end{figure}
\begin{figure}
   \plotone{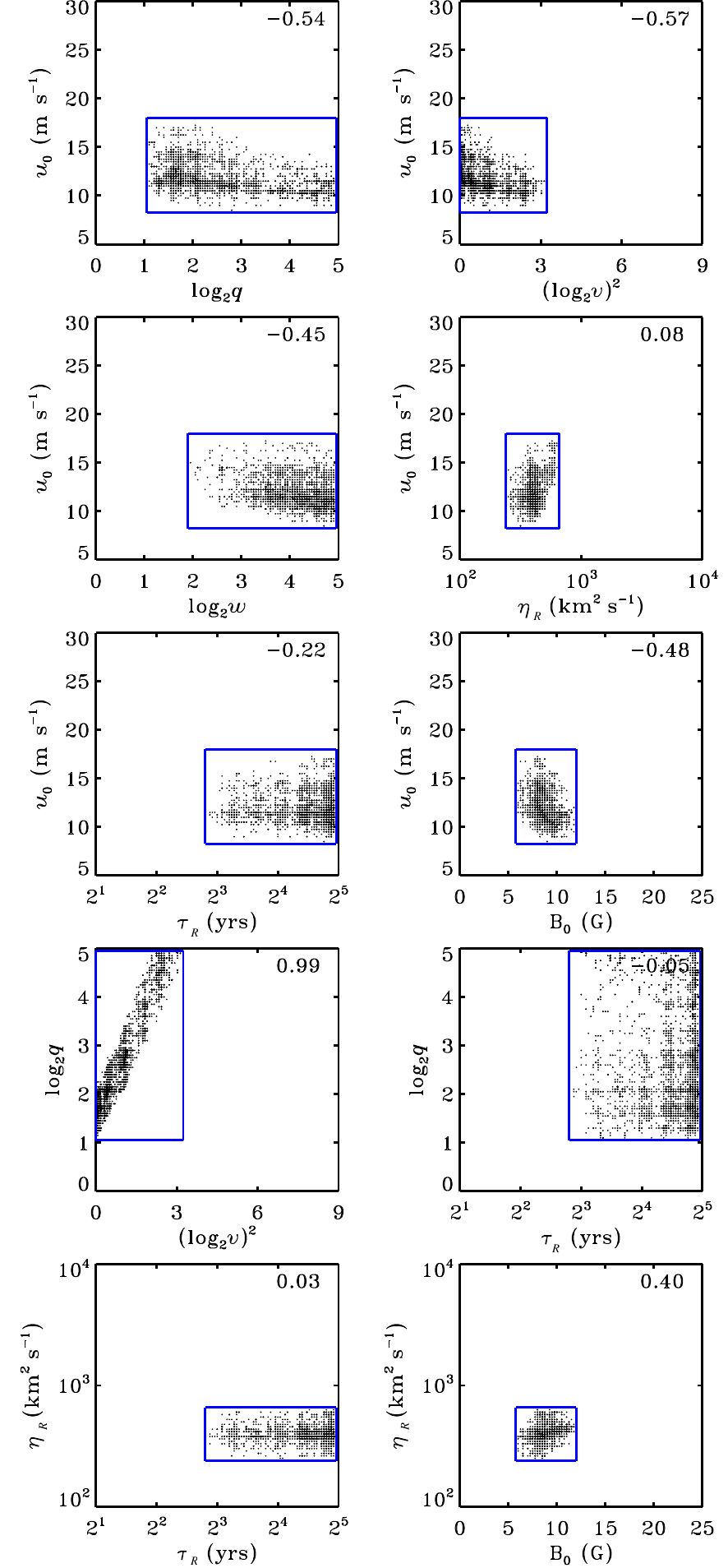}
   \caption{Distribution of $\simeq15000$ acceptable solutions extracted 
   from Figure~\ref{f_fitland}, in ten two-dimensional cuts of the 
   seven-dimensional parameter space.
   Parameter $(\log_2 v)^2$ was plotted instead of $\log_2 v$ for a
   better visualization. 
   Blue boxes delimit those solutions in terms of acceptable intervals 
   for each parameter.
   Also shown are the corresponding the Pearson's linear correlation 
   coefficients.}
   \label{f_fitcorrel}
\end{figure}
Figure~\ref{f_fitland} illustrates the value of criterion $\chi^{-2}$ as 
a function of each parameter value, for a set of $144000$ solutions 
obtained from six independent W21-7 optimizations (different seed 
populations), $500$ generations each, $48$ individuals per generation.
In all six cases, the fitness reaches the same maximum value 
$\chi^{-2}_\text{max}=0.45$. 
Unfortunately, all parameters do not end up constrained equally tightly. 
As a first estimate of fitting errors, we take a look at intervals
for which the maximum value of $\chi^{-2}\geq0.445$.
The corresponding parameter ranges are indicated by the thick horizontal
line segments on each panel of Figure~\ref{f_fitland}. 
We succeed in obtaining narrow optimal peaks for three parameters 
($u_0$, $\eta_R$ and $B_0$), but not for the other four parameters 
($q$, $v$, $w$, and $\tau_R$).

In such a complex modeling problem, the optimal solution is only as 
physically meaningful as the goodness-of-fit measure being maximized by 
the {\GA}.
Our adopted fitness measure (Equation~(\ref{eq_fitns})) is physically 
motivated, but nonetheless retains some level of arbitrariness
(e.g., the exact latitudinal boundaries of our ``transport'' regions, 
and equal weight given to each fitness submeasures). 
Clearly, there must exist a value of $\chi^{-2}$ above which solutions 
are physically acceptable, even if not strictly optimal.
An example of such a solution, with $\chi^{-2}=0.42$, is presented in 
Figure~\ref{f_cyc21}c. 
It corresponds to the optimal solution obtained when minimizing only the 
difference between the two maps (maximum $\chi^{-2}_\text{map}$). 
The time--latitude map, unsigned magnetic flux, and axial dipole moment 
look very similar to the optimal solution. 
The main difference lies more in the shape of the flux-migration 
strips, which are slightly too diffuse and thus less distinct from one 
another, leading to smoother curves for the integrated field in the two 
transport regions (thin blue line in Figure~\ref{f_cyc21}g and h). 
These differences appear significant enough to understand that such a 
solution is not as good as the optimal one, but still at the limit of 
acceptability in terms of observed global features.

\begin{figure}
   \plotone{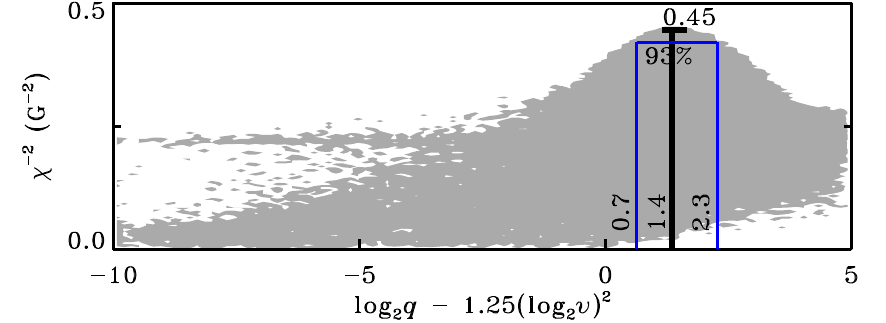}
   \caption{Distribution of the $144000$ solutions tested, as in 
   Figure~\ref{f_fitland}, but now in terms of a combination of 
   parameters $q$ and $v$.}
   \label{f_fitlandqv}
\end{figure}

We go one step further and examine the properties of all solutions
produced by the {\GA} that are characterized by a fitness larger
than $93\%$ of the optimal fitness $\chi^{-2}_\text{max}$.
Among the $144000$ solutions presented in Figure~\ref{f_fitland}, less 
than $15000$ satisfy this criterion, arising from various combinations 
of parameters inside the corresponding intervals (as delineated by the 
thin blue lines in the figure). 
The opposite is not true, however: many combinations of parameters 
inside these intervals still lead to inappropriate solutions that lie 
below the $93\%$ line. 
To exclude the unacceptable solutions and assess how the acceptable 
ones behave inside these intervals,
we explore the shape of the seven-dimensional parameter-space landscape, 
proceeding by pairs of parameters. 
Figure~\ref{f_fitcorrel} illustrates ten representative cuts, 
among the 21 possible combinations. 
If all combinations of two given parameters were good enough, the 
corresponding rectangle would be filled with 
acceptable solutions.
On the other hand, any empty region of a given rectangle indicates that 
the corresponding combination of parameters is not acceptable. 
In particular, the obvious trend observed between $\log_2 q$ and 
$(\log_2 v)^2$, with a correlation coefficient of $0.99$, indicates a 
close linear dependence between the two parameters (of slope $1.25$), 
and that any other combination of the two parameters should be avoided. 
Replotting the distribution of original solutions in terms of 
$\log_2 q - 1.25 (\log_2 v)^2$ instead of $\log_2 q$ 
successfully gives rise to a fourth, well-constrained parameter 
(Figure~\ref{f_fitlandqv}). 

\subsubsection{Parameter $B_0$}

The preceding set of simulations were run using 
profile~(\ref{eq_initcnd1}) as an initial condition.
The bottom panel of Figure~\ref{f_fitland} shows an amplitude converging 
quite properly to $B_0=\unit{8.5\pm\substack{3.5\\2.5}}{\gauss}$. 
This initial state corresponds to an unsigned magnetic flux 
$F_\text{sim}\simeq\unit{6\times10^{22}}{\maxwell}$ (with negligible 
signed flux $\Phi_\text{sim}\simeq\unit{10^{15}}{\maxwell}$) during 
1976 June, to be compared with the improbable unsigned flux 
$F_\text{dat}\simeq\unit{10^{23}}{\maxwell}$ observed for the same 
period, but unfortunately unbalanced by 
$\Phi_\text{dat}\simeq\unit{5\times10^{22}}{\maxwell}$ 
(see Figure~\ref{f_cyc21}e). 
Independent testings of initial condition~(\ref{eq_initcnd2}) have also 
shown the solutions to converge to a maximum fitness 
$\chi^{-2}_\text{max}=0.45$, 
with almost identical results obtained for each parameter, except for 
$B_0^\ast\simeq\unit{6.6}{\gauss}$. 
This lower value for the polar magnetic field amplitude was to be 
expected considering the flatter profile near the poles and corresponds, 
in fact, to the same unsigned flux $F_\text{sim}$. 
As a result, this means that our optimization does not allow us to 
distinguish between initial conditions~(\ref{eq_initcnd1}) and 
(\ref{eq_initcnd2}), but that we can be confident that the net 
unsigned magnetic flux was closer to $\unit{6\times10^{22}}{\maxwell}$ 
than $\unit{10^{23}}{\maxwell}$ during early cycle 21. 
As for the remainder of our analysis, even if 
profile~(\ref{eq_initcnd2}) seems in closer agreement with the 
latitudinal distribution observed at other cycle minima, we choose to 
stick to simplicity with the more conventional 
profile~(\ref{eq_initcnd1}). 
And since the only original purpose of optimizing $B_0$ was to 
ensure a suitable initial state for the optimization of other 
parameters, we go on with our analysis assuming profile 
$B_R(\theta,\phi,t_0) = \unit{8.5}{\gauss}\abs{\cos\theta}^7 \cos\theta$ 
as being representative of the solar photosphere around 1976 June. 
This value of $B_0=\unit{8.5}{\gauss}$ also ensures that maximum 
intervals are still considered for parameters $u_0$ and $\eta_R$, as 
illustrated in the corresponding plots of Figure~\ref{f_fitcorrel}.

\subsubsection{Parameter $\tau_R$}

Figure~\ref{f_fitland} (sixth panel) reveals a very smooth distribution 
of solutions in terms of parameter $\tau_R$, unfortunately without any 
peak inside the explored domain. 
A wide variety of solutions lie above the $93\%\chi^{-2}_\text{max}$ 
line, meaning that various combinations of parameters with $\tau_R$ 
between $7$ and $\unit{32}{\years}$ can lead to acceptable solutions. 
In fact, this parameter was expected to be difficult to 
constrain from the optimization of a single solar cycle,
though \citet{Yeates2014} finds a value of $\unit{10}{\years}$ to better
reproduce the evolution of cycle 23 and early cycle 24. 
We must recall that parameter $\tau_R$, with a value of 
$\unit{5-10}{\years}$, was found by \citet{Schrijver2002} to be required 
in the equation of {\SFT} to allow some exponential 
decay of accumulated flux,
and thus allow polarity reversals even when two 
subsequent cycles have markedly different amplitudes.
Since our optimization process appears unable to constrain the 
parameter, we opt to use $\tau_R=\unit{32}{\years}$ for the remainder 
of the present analysis, which is approximately equivalent to simply 
doing away with the linear term $-B_R/\tau_R$ in 
Equation~(\ref{eq_surftrans}).


\subsubsection{Meridional circulation profile}

\begin{figure}
   \plotone{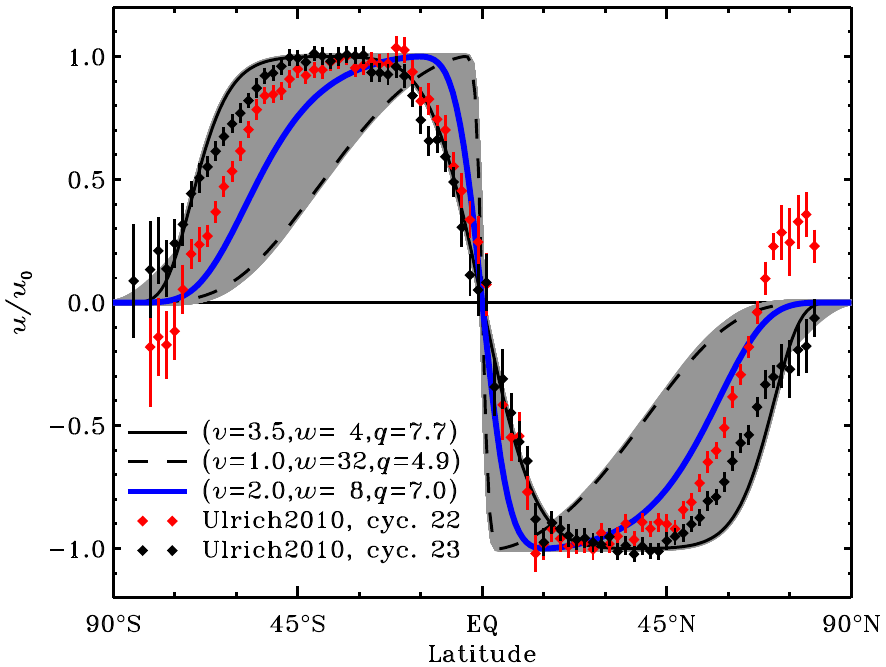}
   \caption{Our optimal meridional circulation surface profile (thick 
   blue curve), with shaded area illustrating all possible curves within 
   the error bars listed in Table~\ref{t_param}.
   Thin continuous and dashed black lines indicate some sample extreme 
   solutions.
   Red and black dots and error bars illustrate average measurements for 
   cycles 22 and 23, respectively, as obtained by \citet{Ulrich2010}.
   All curves are normalized with respect to maximum flow speed $u_0$.}
   \label{f_mercircfinal}
\end{figure}

The second, third, and fourth panels of Figure~\ref{f_fitland} also 
reveal a wide variety of solutions above $93\%\chi^{-2}_\text{max}$ for 
parameters $q$, $v$, and $w$ ($q\in[2,32]$, $v\in[1.0,3.5]$, 
$w\in[4,32]$), suggesting that our optimization is rather bad at 
constraining our sophisticated latitudinal profile 
(Equation~(\ref{eq_mercirc})).
Fortunately, the strong correlation observed between $\log_2q$ and 
$(\log_2v)^2$ (Figure~\ref{f_fitcorrel}) and the fairly narrow peak 
obtained for $\log_2 q - 1.25 (\log_2 v)^2$ (Figure~\ref{f_fitlandqv}), 
reveal that one of those three parameters can be, in fact, rather well 
constrained. 
The result is a parameter $q$ that is restricted to 
$2.8\pm\substack{2.0\\1.1}$ when $v=1$ 
and up to ${\gtrsim}\,30$ when $v=3.5$.
Generally formulated, this gives 
$q={\left(2.8\pm\substack{2.0\\1.1}\right)}\cdot{2^{1.25(\log_2 v)^2}}$. 
The effect is a surface flow that tends to drop to zero before reaching 
the poles, usually between $\unit{70}{\degree}$ and $\unit{80}{\degree}$ 
latitude, regardless of the value of $v$ 
(see Figure~\ref{f_mercircfinal}). 
This allows the build-up of magnetic polar caps that are not 
too confined near the poles. 
Such high-latitude behavior is compatible with meridional flow profiles 
by \cite{vanBalle1998} and \cite{Wang2002b} (see Figure~\ref{f_mercirc}).

Furthermore, the interval obtained for parameter $w\in[4,32]$ is not as 
unconstraining as it may appear. 
It corresponds to a latitudinal flow that reaches its peak speed between 
$\unit{4}{\degree}$ and $\unit{25}{\degree}$ latitude, which is actually 
a noteworthy result since it excludes many typical profiles 
used in the literature that tend to peak at or near mid--latitudes 
\citep[e.g.,][]{Komm1993,Dikpati1999}, in particular the profile by 
\citet{vanBalle1998} used earlier as a reference case. 
This quick rise of the flow speed between the equator and 
$\unit{4}{\degree}$--$\unit{25}{\degree}$ latitude seems to be required 
to prevent too 
large a cross-hemispheric cancellation of the {\BMR}s' western flux, and 
thus too large a net flux to be transported to higher latitudes, as 
happened in the reference case. 

The dominant uncertainty that remains in our surface circulation 
profile concerns parameter $v$, whose solutions between $1$ and $3.5$ 
correspond to a surface flow peak that has respectively no extent in 
latitude (rapid rise from the equator to 
$\unit{4}{\degree}$--$\unit{25}{\degree}$ 
latitude and immediate decrease) or a width up to $\unit{45}{\degree}$ 
(rapid rise from the equator to 
$\unit{4}{\degree}$--$\unit{25}{\degree}$ latitude, 
followed by a plateau up to $\unit{49}{\degree}$--$\unit{55}{\degree}$).
This result is obviously imprecise since it constrains rather poorly 
the flow speed between $\simeq\unit{25}{\degree}$ and 
$\simeq\unit{70}{\degree}$ latitude (see Figure~\ref{f_mercircfinal}).
This is quite surprising since these latitudes harbor the ``transport 
regions'' where flux-migration strips build up.

The problem is partly alleviated by the uncertainty in the maximum flow 
speed $u_0$. 
As can be seen in the top right panel of Figure~\ref{f_fitcorrel}, 
the allowed interval for $u_0$ varies from $\unit{9-17}{\mps}$ when 
$v=1$ to $\unit{10-12}{\mps}$ when $v\simeq3.5$. 
The resulting flow speed in the transport regions 
(latitudes $\unit{34}{\degree}$--$\unit{51}{\degree}$) ends up being 
more dependent on $u_0$ than $v$, with values of 
$\unit{3-14}{\mps}$ when $v=1$ to $\unit{9-12}{\mps}$ when $v\simeq3.5$. 
The uncertainty obviously remains substantial, and since it is unlikely 
that the latitudinal flow speed does not influence the shape, 
especially the inclination, of flux-migration strips, this result means 
either that the calculation of average magnetic field 
$\langle B_R \rangle^\text{T1,T2}$ in the transport regions is not a 
sufficiently restrictive way to characterize these shapes 
or that the discrepancies between the observed and simulated curves 
of $\langle B_R \rangle^\text{T1,T2}$ are too large to allow a selective 
comparison.
Nonetheless, the mid--latitude features observed in Figure~\ref{f_cyc21}c 
and \ref{f_cyc21}d do fit better than those of the reference case 
(Figure~\ref{f_cyc21}b).
The solution must therefore come from a delicate equilibrium between 
advection and diffusion.

Figure~\ref{f_mercircfinal} plots the whole variety of acceptable 
latitudinal profiles described above in the form of a shaded area.
Also superimposed on the figure are average Doppler measurements provided 
by \citet{Ulrich2010} for cycles 22 and 23.
Apart from some high-latitude equatorward flow observed for cycle 22, 
which we deliberately
opt to ignore, all measurements fit quite nicely inside the optimal 
shaded area.
More specifically, measurements for cycle 22 can be well approximated 
below
$\simeq\unit{65}{\degree}$ latitude by a $(v=2.7,w=4,q=11)$ curve, 
and cycle 23's pattern up to $\simeq\unit{75}{\degree}$ latitude by a 
$(v=2.7, w=4, q=5)$ curve.
In terms of amplitude, cycle 22's smoothed trend peaks near 
$\unit{14-15}{\mps}$ while that of cycle 23 peaks closer to 
$\unit{16}{\mps}$.
In both cases, the two hemispheres are not perfectly symmetric.
All these values fit adequately inside, or at the limit of, the optimal 
intervals listed in Table~\ref{t_param}.

The completion of our analysis now requires the selection of 
a final representative profile for cycle 21.
To remain independent from direct meridional flow measurements, we 
instead choose some reasonable profile that lies near the middle of the 
optimal region.
A value of $w=8$ seems reasonable in terms of peak flow latitude 
($\unit{15}{\degree}$) and near-equator latitudinal gradient, and 
prevents numerical instabilities that could occur at low spatial 
resolution with higher values of $w$. 
With parameter $v=2$, the fast low-latitude flow slows down only 
slightly up to $\simeq\unit{45}{\degree}$ latitude, an apparently good 
compromise between a purely peaked profile and a broad sustained 
plateau. 
This leaves the allowed interval $[4,11]$ for parameter $q$, with a peak 
at $q=7$, for a final drop to zero speed near 
$\unit{70}{\degree}$--$\unit{75}{\degree}$ latitude.
This final profile is shown in blue in Figure~\ref{f_mercircfinal}.

\subsubsection{Maximum flow amplitude and magnetic diffusivity}

\begin{figure}
   \plotone{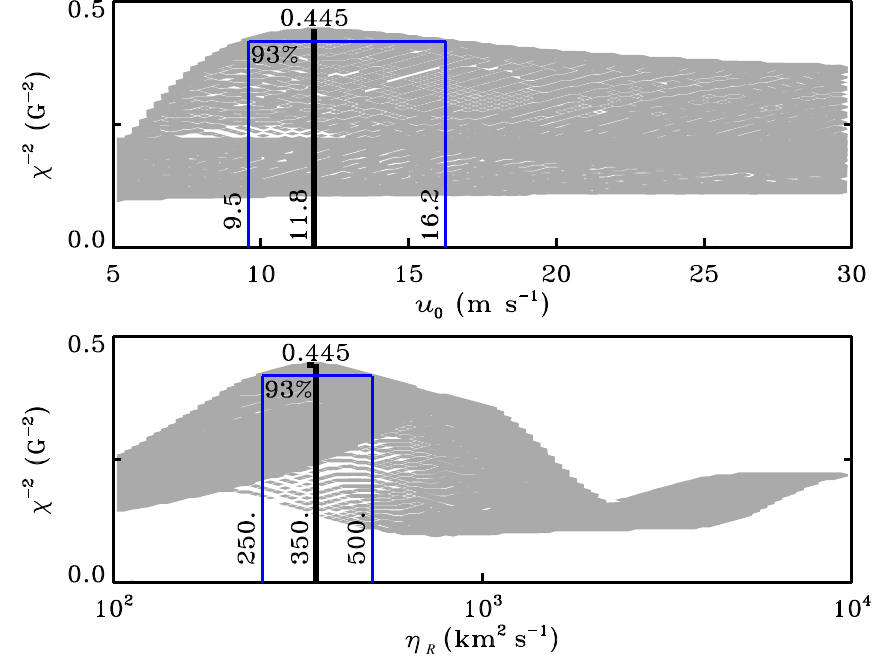}
   \plotone{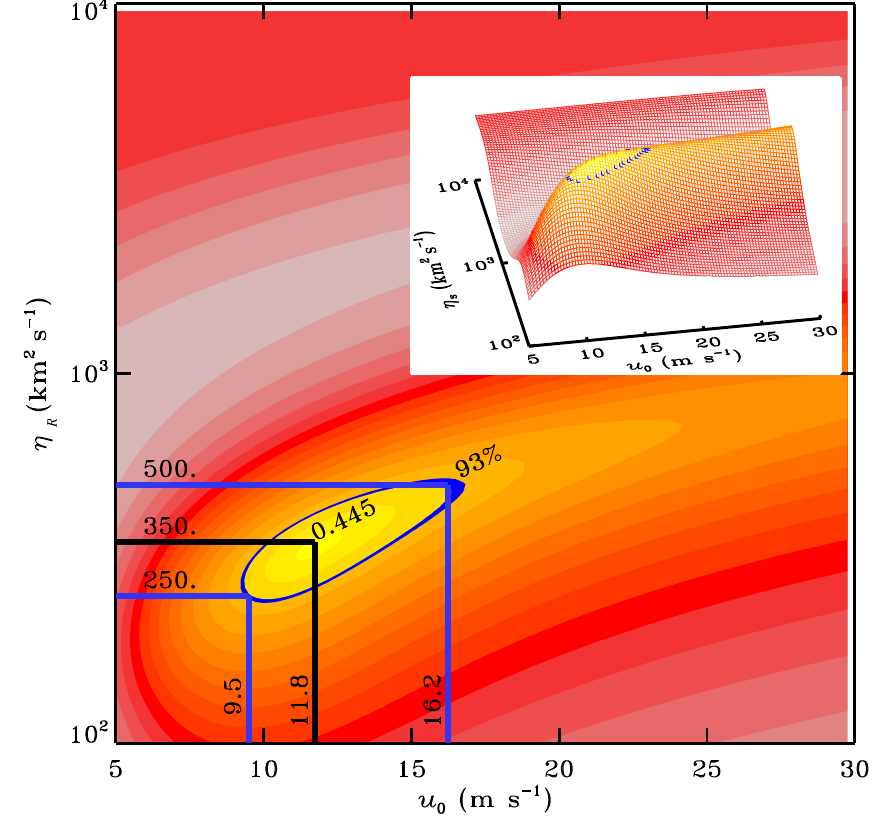}
   \caption{Fitness $\chi^{-2}$ as a function of parameters $u_0$ and 
   $\eta_R$, from optimization W21-2.
   Thick black lines indicate the position of the best solution.
   Thin vertical blue lines indicate the parameter values where fitness 
   reaches $93\%\chi^{-2}_\text{max}$, such that any solution above the 
   blue line or inside the blue ring is considered acceptable.}
   \label{f_fitland2d}
\end{figure}

Although Figure~\ref{f_fitland} shows clear optimal peaks for parameters 
$u_0$ and $\eta_R$, a wide variety of solutions in the 
intervals $u_0\in\unit{[8,18]}{\mps}$ and 
$\eta_R\in\unit{[240,660]}{\kmmps}$ 
do not reach the $93\%\chi^{-2}_\text{max}$ line. 
Considering the various interdependences illustrated in 
Figure~\ref{f_fitcorrel}, we opt to
perform a last cycle 21 optimization, called W21-2,
using the fixed meridional circulation profile 
chosen above and only $u_0$ and $\eta_R$ as free parameters. 
We cover the whole domain with $10000$ solutions and obtain the 
two-dimensional landscape illustrated in Figure~\ref{f_fitland2d}.

The maximum fitness obtained lies $1\%$ lower than the original 
$\chi^{-2}_\text{max}=0.45$, due to the use of a slightly suboptimal 
$w=8$ in our final surface flow profile. 
Nonetheless, a clear peak rises above the 
$93\%\chi^{-2}_\text{max}$ ring, within the intervals 
$u_0\in\unit{[10,16]}{\mps}$ and $\eta_R\in\unit{[250,500]}{\kmmps}$. 
These values both roughly correspond to those found in the literature, 
though $\eta_R$ does not include the $\unit{600}{\kmmps}$ used in the 
reference case and typical of many studies 
\citep[e.g.][]{Wang1989,Mackay2002-2}. 
The acceptable combinations of the two parameters form an elongated 
ridge, already noticeable in the equivalent plot of 
Figure~\ref{f_fitcorrel}, now with a much higher linear correlation of 
$0.81$. 
This positive interdependence illustrates the aforementioned
delicate balance between advection and diffusion.
In fact, a faster latitudinal flow gives less time for cancellation to 
occur between opposite polarities of the {\BMR}s before they reach the 
poles, thus requiring a higher diffusivity. 
Similarly, mid--latitude flux strips will keep the same width with a 
higher diffusivity, provided the flux is transported more quickly. 
This balance is also required at the poles, where a stronger flow would 
squeeze the magnetic cap to higher latitudes if it were not for a higher 
diffusivity.

The best linear fit to this final restrained region gives a slope of 
$0.037$, such that parameters $u_0$ and $\log\eta_R{-}0.037u_0$ are 
nearly independent. 
As a final numerical constraint to those two parameters, we have 
$u_0=\unit{12\pm\substack{4\\2}}{\mps}$ and 
$\eta_R=\unit{(350\pm70)\cdot10^{0.037(u_0-12)}}{\kmmps}$, with an 
overall range of $\unit{[250,500]}{\kmmps}$ for $\eta_R$.
These intervals correspond to a magnetic Reynolds number
$R_\text{m}=24\pm\substack{6\\5}$.

These numerical values overlap with results obtained in analyses of 
advection--diffusion-based SFT simulations 
by \citet{Wang1989} and \citet{Wang1991}, who found values of
$\eta_R=\unit{600\pm200}{\kmmps}$ and $u_0=\unit{10\pm3}{\mps}$ to
to be required for reproducing the evolution of polar field strengths, 
dipole strengths, and large-scale open magnetic flux, as well as values
of $\eta_R=\unit{600}{\kmmps}$ and $u_0=\unit{11}{\mps}$ used by 
\citet{Baumann2004} as a reference case.
However, their latitudinal flow profiles are to be excluded by the 
interval of optimal profiles described above.
On the other hand, \citet{Wang2002b} found 
$\eta_R\simeq\unit{500}{\kmmps}$, with a surface flow profile   
(see Figure~\ref{f_mercirc}) that fits the optimal constraints detailed 
in Table~\ref{t_param}, but an amplitude $u_0\simeq\unit{20-25}{\mps}$
that is definitely outside of our fitted boundaries.
Alternatively, random-walk-based surface flux evolution models tend
to lead to smaller diffusion coefficients (see, e.g., 
\citet{Schrijver2001} who found $\eta_R\simeq\unit{300}{\kmmps}$,
and \citet{Thibault2014} who used $\eta_R\simeq\unit{416}{\kmmps}$)
Furthermore, for the lower range of $u_0$, our result for $\eta_R$
overlaps with indirect measurements by, e.g., \citet{Mosher1977} and 
\citet{Komm1995} who obtained values in the range 
$\simeq\unit{100-300}{\kmmps}$
(see also \citealt[Table 6.2]{Schrijver2000}, for a compilation of 
published diffusion coefficients).
Last, but certainly not least,
our optimal meridional flow amplitude is in agreement both with 
the tracking of surface magnetic features by \citet{Komm1993} 
($u_0\simeq\unit{13.2}{\mps}$) for cycle 21 and
with Doppler determinations of \citet{Ulrich2010}
($u_0\simeq\unit{14-16}{\mps}$) for cycles 22 and 23.


\subsection{Variable meridional flow}

In \S~\ref{s_optsol} we explained how even our optimal solution does not 
perfectly reproduce some of the polar surges and mid--latitude flux 
strips observed for cycle 21. While we have already explored in detail 
the possible latitudinal variations of the meridional flow speed, one 
explanation for the discrepancies could come from some temporal 
variability.
Such time dependence of the flow is in fact observed 
(see, e.g., \citealt{Ulrich2010}), both in amplitude and shape.

We use our optimization procedure to test for possible improvements
to the best-fit solution by allowing for temporal variations of
the meridional flow amplitude.
We opt for a piecewise-continuous representation of the flow
parameter $u_0$, by successively separating the cycle into $M=2$, $M=4$, 
and $M=8$ contiguous segments of equal duration, each such interval 
having its own value $u_{0,m}$. 
As in the previous W21-2 analysis, we keep the initial condition 
($B_0$), decay time ($\tau_R$), and meridional flow profile (parameters 
$q$, $v$, and $w$) fixed, while optimizing the two, four, or eight 
values for $u_{0,m}$ along with the supergranular diffusivity $\eta_R$.

The addition of more temporal intervals successively improves the 
overall fitness $\chi^{-2}$, by up to $3\%$ ($\chi^{-2}\leq0.46$).
As expected, the improvement is mostly noticeable in sub-criteria 
$\chi^{-2}_\text{T1}$ and $\chi^{-2}_\text{T2}$, which measure the shape 
of mid--latitude inclined strips. However,
the additional degrees of freedom in the optimization 
process also worsen parameter degeneracies; the values of 
$u_{0,m}$ end up simply unconstrained for most of the temporal 
intervals.
These optimization experiments do indicate that the value of $u_0$ in 
the first half of the cycle is most critical, because it is constrained 
adequately, but no robust intra-cycle 
temporal trend can be extracted from the fitting.
Note, however, that in all cases the resulting optimal interval for 
parameter $\eta_R$ remains essentially the same as obtained earlier from
the W21-2 optimization.
The availability of uniform emergence databases for other activity 
cycles may allow, in the future, an extension of the present study over
multiple solar cycles, and perhaps the extraction of statistically 
significant temporal dependences in the meridional flow amplitude.


\section{Emergence-related variability}



Extensive studies (e.g., \citealt{Cameron2014}, and references therein)
have shown that details of the emergence of individual BMRs can
have a strong impact on global cycle properties; this is hardly
surprising considering that a single large BMR contains
about as much magnetic flux as the polar caps at cycle minimum.
In order to quantify the effects of the specific realization of 
bipolar emergences during a given solar cycle, as compared to the 
overall statistics of those emergences, we develop
a Monte Carlo procedure generating synthetic databases of {\BMR}s, 
respecting the statistical properties characterizing latitudinal and 
area distributions of emergences as a function of solar cycle amplitude, 
as established on the basis of the temporally extended photographic 
records from the Royal Greenwich Observatory (RGO), the US Air Force 
(USAF), and the National Oceanic and Atmospheric Administration (NOAA).
The required additional magnetic properties are synthesized using 
the statistical distributions of magnetic fluxes, angular separations, 
and east--west tilts characterizing {\WS}'s database.
Appendix~\ref{a_datab} describes in detail the required analyses.
The end product is a Monte Carlo engine that can generate statistically 
independent realizations of {\BMR} emergences using as the only input 
the monthly value of the International Sunspot Number and the amplitude 
and length of the activity cycles we aim to model.

Accordingly, we compare the results of the {\SFT} simulation based on 
synthetic realizations of cycle 21, with those previously obtained with 
the real emergences compiled by {\WS}.
We generate $\simeq1000$ such independent synthetic realizations, using 
the optimal model parameter values previously obtained for cycle 21 
(Table~\ref{t_param}) to compute the resulting surface magnetic flux 
evolution.
Only three ($0.3\%$) among the thousand simulations based on synthetic 
emergences lead to synthetic magnetograms resembling observations
sufficiently to reach our former acceptable limit of
$93\%\chi^{-2}_\text{max}=0.42$.

\begin{figure}
   \plotone{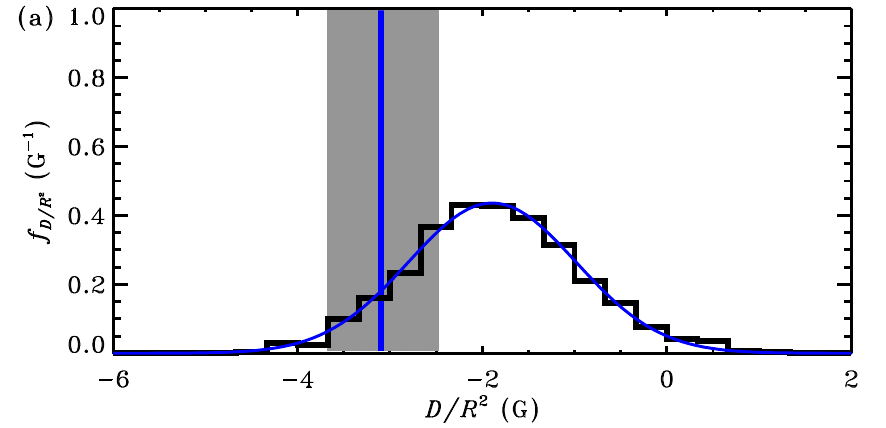}
   \plotone{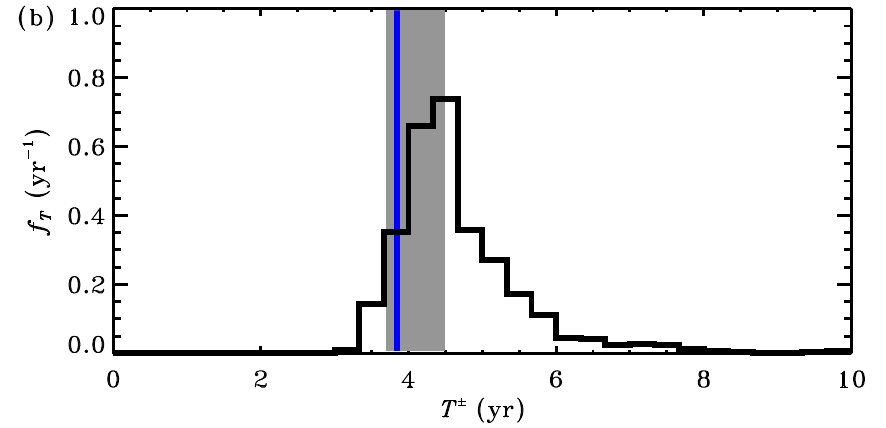}
   \plotone{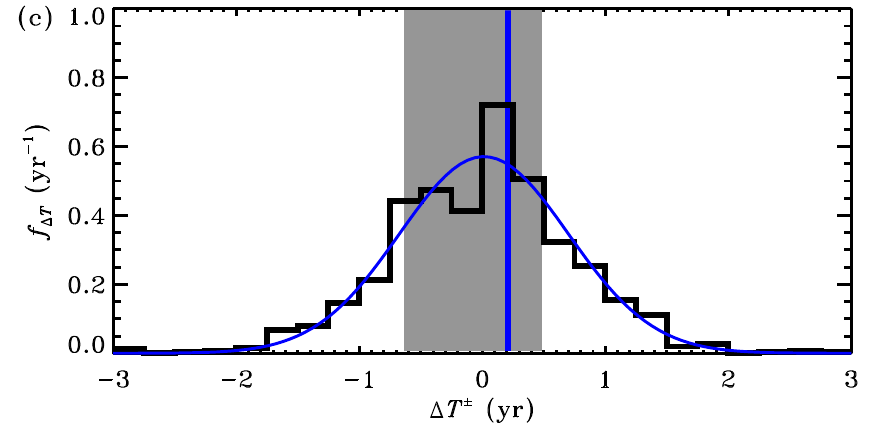}
   \caption{PDFs (thick black histograms) of
   (a) axial dipole moment at the end of cycle 21,
   (b) time required for polarity reversal, since the beginning of cycle 
   21, and
   (c) delay between polarity reversals in the northern and southern 
   hemisphere,
   for $\simeq1000$ databases of synthetic emergences for cycle 21
   generated from independent random sequences.
   Superimposed on (a) and (c) are best gaussian fits (thin blue lines).
   Each plot also shows the corresponding values extracted from our 
   optimal solution
   (vertical blue line; see Figure~\ref{f_cyc21}f), as well as intervals
   (shaded gray area) covered by acceptable solutions in terms of
   criterion~(\ref{eq_fitns}) (see Table~\ref{t_param}).}
   \label{f_synth}
\end{figure}

Criterion $\chi$ (Equation~(\ref{eq_fitns})) was elaborated with the aim of 
constraining the {\SFT} simulation to reproduce the details of 
synoptic magnetograms.
We should perhaps not count on a synthetic realization of cycle 21 to 
perform very well in this respect, but it is still reasonable to 
expect some of the overall trends and global cycle properties
to be reproduced in a probabilistic sense.
We consider here the following three quantities:
the axial dipole moment at the end of cycle 21 ($D^\ast_\text{end}$),
the timing of polarity reversal of the axial dipole moment ($T^\pm$),
and the delay between polarity reversals in the northern and southern 
hemispheres ($\Delta T^\pm$).
In Figure~\ref{f_synth}, we plot their probability density functions 
({\PDF}), built from the thousand synthetic 
realizations generated earlier.
In parallel, we also compute the same three quantities, for a few 
hundred simulations using the {\WS} emergence database, but now with 
the parameter values for the {\SFT} model extracted randomly in the 
intervals given in Table~\ref{t_param}.
The resulting ranges obtained for 
$D^\ast_\text{end}$, $T^\pm$, and $\Delta T^\pm$ are indicated
by the shaded areas in Figure~\ref{f_synth}. 

The {\PDF} of $D^\ast_\text{end}$ based on synthetic emergences is 
Gaussian-shaped, with a peak probability at $\unit{-1.9}{\gauss}$ and 
a standard deviation of $\unit{0.9}{\gauss}$.
This distribution clearly overlaps with the range of optimal 
cycle-21-like solutions, which confirms that the synthetic databases are 
able to reproduce the observed axial dipole moment.
However, most realizations of synthetic emergences end up 
building a weaker axial dipole moment than the
$D^\ast_\text{end}=\unit{-3.1\pm0.6}{\gauss}$ covered by acceptable 
solutions for the real cycle 21.

The distribution for the time $T^\pm$ of polarity reversals in our 
synthetic cycle 21 peaks near $\unit{4.7}{\years}$, 
with a standard deviation of $\simeq\unit{1.0}{\years}$, and an 
asymmetric shape suggesting that polarity reversals are more difficult 
to hasten than to delay.
The optimal solutions
($T^\pm=\unit{3.8\pm\substack{0.6\\0.1}}{\years}$), and presumably the 
real cycle 21, now lie closer to the peak of probability.
The few simulations that show a highly delayed or even no reversal 
($T^\pm\rightarrow\unit{10}{\years}$ in Figure~\ref{f_synth}b) 
correspond to those with a weaker or even positive axial dipole moment 
at cycle minimum ($D^\ast_\text{end}\simeq\unit{0}{\gauss}$ on 
Figure~\ref{f_synth}a).

Finally, the distribution of $\Delta T^\pm$ indicates that hemispheres 
tend to be in phase or nearly so in most simulations, as for the 
observed cycle 21, but still with a standard deviation of 
$\unit{0.7}{\years}$.

We conclude from this exercise that uncertainties in global cycle
characteristics are dominated by the inherent stochasticity of the
flux emergence process, rather than by uncertainties related to
model calibration.
This inherent stochasticity is therefore what is likely to limit the 
predictive capability of any dynamo-based solar cycle prediction 
schemes, highlighting the need for appropriate data assimilation
procedures, and dynamo models suitably designed toward this end.

\section{Conclusions}

We have reported in this paper on the design of the surface component of 
a coupled surface--interior Babcock--Leighton dynamo model of the solar
cycle, including a latitudinally and longitudinally resolved 
representation of the solar photospheric magnetic field.
Specifically, we used a genetic algorithm to evolve a surface magnetic 
flux evolution model providing an optimal representation of a surface 
synoptic magnetogram.
Our procedure is robust, in that it can operate in multimodal parameter 
spaces and escape secondary extrema.
It also returns useful error estimates on all best-fit parameters, and 
allows the identification of any correlations between these parameters.

An essential input to any surface magnetic flux evolution model is the 
characterization of emerging BMRs in the course of an activity cycle:
their time of emergence, latitude and longitude, magnetic flux,
pole separation, and tilt with respect to the east--west direction.
\citet{Wang1989-0} have assembled an appropriate database for cycle 21, 
covering the period 1976 August to 1986 April, which they kindly
made available to us.
The optimization of our magnetic flux transport model was therefore 
carried out using this database, over that same time period.

The optimal solution is characterized by a surface magnetic diffusivity
intrinsically correlated with the amplitude of the surface meridional
flow speed, that is 
$\eta_R=\unit{(350\pm70)\cdot10^{0.037(u_0-12)}}{\kmmps}$ for
$u_0=\unit{12\pm\substack{4\\2}}{\mps}$.
This interval of solutions for $\eta_R$ is in agreement with analyses 
by, e.g., \citet{Wang1989}, \citet{Wang2002b}, \citet{Schrijver2001}, 
\citet{Dikpati2004}, and \citet{Cameron2010}, 
and, for the lower range of $u_0$, is also compatible with indirect
measurements by \citet{Mosher1977} and \citet{Komm1995}.
The meridional flow amplitude is in agreement both with the tracking of 
surface magnetic features by \citet{Komm1993} and
with Doppler determinations of \citet{Ulrich2010}.
The latitudinal dependence of the optimal surface meridional flow 
profile is found to be in good agreement with \citet{Ulrich2010}'s 
measurements, even though these data are not used in the optimization 
process.
This provides an independent validation of our best-fit models.
While the latitudinal profile of the surface meridional flow is not 
entirely constrained by our fitting procedure, it is sufficiently 
limited to exclude a number of latitudinal profiles commonly used in 
extant flux transport models.

Prior modeling work has demonstrated quite clearly that the global
aspects of surface magnetic flux evolution over an activity cycle, in 
particular the timing of the polarity reversals
and strength of the dipole moment at the end of the cycle, are 
sensitively dependent on details of magnetic flux emergence, and in 
particular on the frequency and properties of large BMRs 
emerging close to the equator \citep{Cameron2014}.
In this respect, the observed cycle 21 represents one possible 
realization of an activity cycle.
Using the unified sunspot group database of D. Hathaway as well as the 
aforementioned database of Wang and Sheeley,
we designed a Monte Carlo simulation of BMR emergence in which 
emergence statistics are tailored to reproduce observed statistics.
Our overall procedure is similar to that presented in \citet{Jiang2011a} 
but differs in a number of significant details.
We used this Monte Carlo procedure to generate a large set of synthetic
realizations of cycle 21.
This allowed us to quantify the degree to which global surface magnetic 
flux evolution is impacted by idiosyncrasies of BMR emergences.
The timing of polarity reversals and the associated time delay between 
solar hemispheres are both fairly robust (standard deviations of 
$\unit{1.0}{\years}$ and $\unit{8}{months}$, respectively),
with mean values close to those obtained for the observed cycle 21.
The dipole moment at the end of the cycle, on the other hand, shows 
greater variability, $D^\ast_\text{end}=\unit{-1.9\pm0.9}{\gauss}$, 
with a mean value significantly smaller than that of the observed cycle 
21 ($D^\ast_\text{end}=\unit{-3.1\pm0.6}{\gauss}$ for the set of 
acceptable best-fit models).

In the following paper in this series 
(A. Lemerle \& P. Charbonneau 2015, in preparation),
we couple the calibrated surface transport model discussed herein to a 
kinematic axisymmetric mean-field-like flux transport dynamo model.
With the surface flux model effectively providing a 
Babcock--Leighton-like source term through the upper boundary condition 
on the dynamo model, and the latter providing emergences to feed the 
former, there results a working solar cycle model where the 
stochasticity in surface flux emergence and transport self-consistently 
feeds back into the dynamo loop.
This represents a unique analysis tool toward the understanding of the 
origin of solar cycle fluctuations, as well as a computational framework 
ideally suited for assimilation of magnetographic data toward cycle 
prediction.
\begin{figure*}
   \plotone{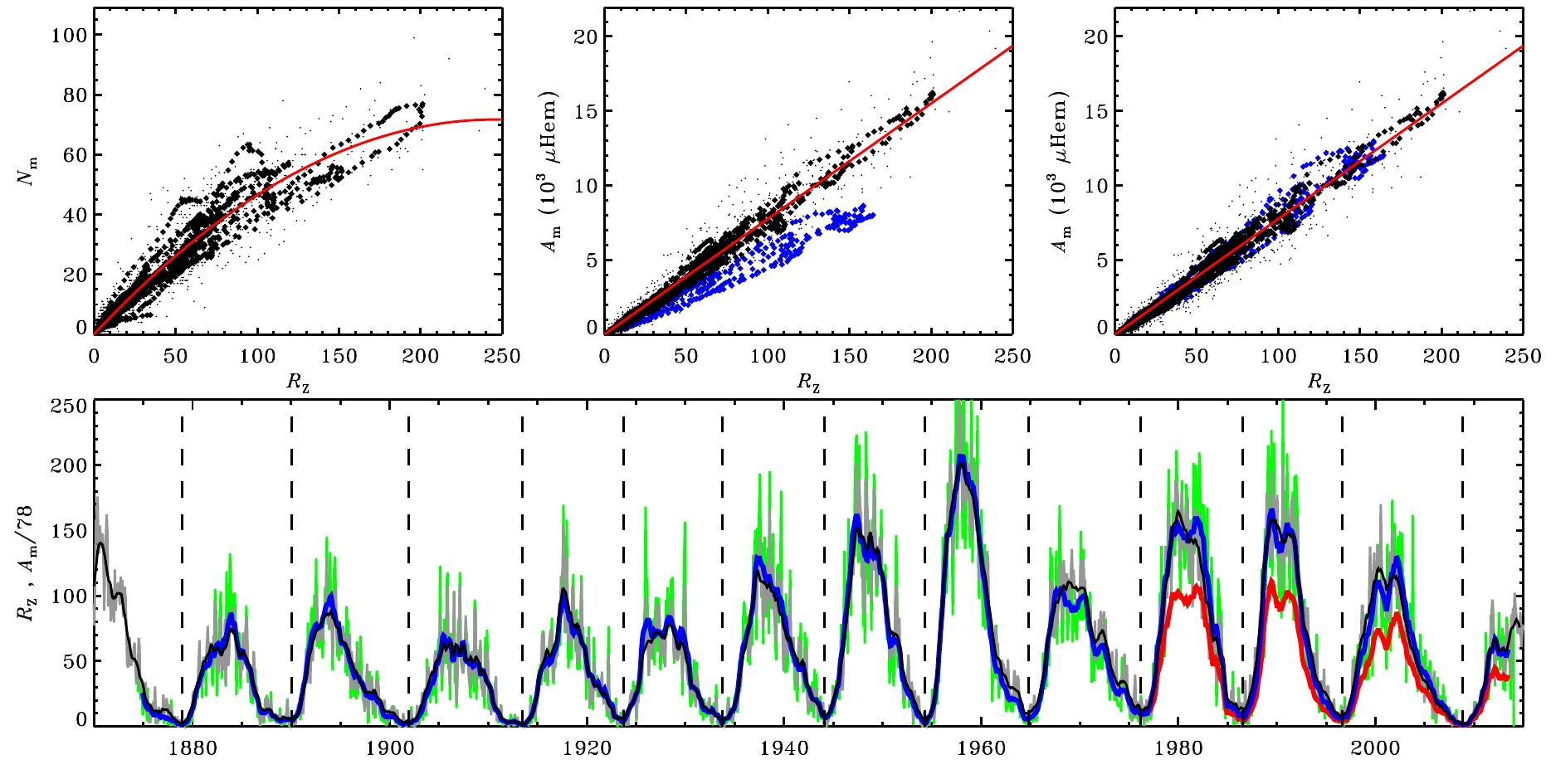}
   \caption{Top left panel: actual number of emergences $N_\text{m}$ 
   during a given month of the RGO data plotted against monthly Sunspot 
   Number $R_\text{Z}$ (small dots); corresponding 13-months running 
   means (thick dots); and best quadratic fit to these last (red line). 
   Center panel: total area $A_\text{m}$ of all emergences during a 
   given month of the RGO data plotted against $R_\text{Z}$ (small black 
   dots); corresponding 13-months running means (thick black dots); best 
   linear fit to these last (red line); and 13-months running mean of 
   USAF-NOAA $A_\text{m}$ plotted against the 13-months running mean of 
   $R_\text{Z}$ (thick blue dots).
   Top right panel: same as center panel, but with USAF-NOAA areas 
   multiplied by $1.5$.
   Bottom panel: temporal evolution of $R_\text{Z}$ (thin grey), its 
   13-months running mean (thin black), RGO $A_\text{m}$ (1876-1976) 
   followed by USAF-NOAA areas (1977-2013) corrected by a factor $1.5$ 
   (thin green), the corresponding 13-months running mean (thick blue), 
   and the uncorrected USAF-NOAA areas (thick red).
   Vertical dashed lines indicate the times of solar cycle minima.}
   \label{f_rzamcor}
\end{figure*}

%


\acknowledgments
We wish to thank
Yi-Ming Wang and Neil R. Sheeley, Jr. for kindy providing us with their 
comprehensive database of bipolar emergences for cycle 21,  
David H. Hathaway for his magnetic butterfly diagram data, 
and Roger Ulrich for his compilation of latitudinal flow measurements 
and error estimates.
This research was funded by
a doctoral research scholarship of the Fonds de Recherche du Qu\'ebec 
Nature et Technologies (A. L.),
and a summer research scholarship (A. C.-D.)
and the Discovery Grant Program (P. C.)
of the Natural Science and Engineering Research Council of Canada.
Calculations were performed on Calcul Qu\'ebec's computing facilities,
a member of Compute Canada consortium.




\appendix

\section{Synthetic database}\label{a_datab}

The detailed analyses presented in \S~\ref{s_cyc21} were made possible 
by the availability of the {\WS} emergence database, which provides the 
input required for our {\SFT} simulation: times of emergence, 
heliographic positions, tilts, separations, magnetic fluxes, and 
polarity.
To the best of our knowledge, no similarly comprehensive emergence
database is currently available for other sunspot cycles
(but do see \citealt{Yeates2014}).

A workaround lies in building a synthetic database grounded on observed 
statistics, rather than real individual magnetic emergences.
A suitable long-term record of daily observations of sunspots groups has 
been assembled by David Hathaway, at the Marshall Space Flight 
Center\footnote{\url{http://solarscience.msfc.nasa.gov/greenwch.shtml}}, 
combining an old record (1874-1976) from the RGO with more recent data from the USAF and the NOAA.
Following in essence the approach described in \citet{Jiang2011a},
we start our analysis by an exploration of this non-magnetic database 
to extract the statistical behavior of the position and umbral area 
of sunspot groups as a function of time, and link it to the more
temporally extended monthly International Sunspot Number.
To consider each group only once, we extract its properties when it 
reaches maximum area.
We finally arrive at a synthetic magnetic database by statistically 
filling the gap between RGO--USAF--NOAA and {\WS} databases.

\subsection{Number of Emergences}

Our aim is to construct a synthetic database of sunspot groups, 
and hence {\BMR}s, generated solely from the smoothed monthly value of 
the International Sunspot Number ($R_\text{Z}$), as well as the 
corresponding $n\text{th}$ cycle amplitude, given by the maximum 
value of $R_\text{Z}$ inside the cycle ($R_{\text{Z}~n}^\text{max}$), 
and length ($L_n$), where cycles are delimited by times of minimum 
activity as obtained from the average of three solar indexes by 
\citet{Hathaway2010}
(see vertical dashed lines in Figure~\ref{f_rzamcor}).
A first step would be to determine the number of sunspot groups 
($N_\text{m}$), and indirectly the number of {\BMR}s, to emerge during a 
given month.
By definition, $R_\text{Z}$ is related nonlinearly to $N_\text{m}$, and 
a direct linear correlation between the two quantities is thus unlikely 
to be good.
Instead, we find a much better linear correlation between $R_\text{Z}$ 
and the monthly total area $A_\text{m}$ of those $N_\text{m}$ emerged 
groups.
The upper middle and right panels of Figure~\ref{f_rzamcor} show the 
good linear correlation that exists between RGO's $A_\text{m}$ and 
$R_\text{Z}$ ($r=0.96$), especially when considering their respective 
13 month running mean ($r=0.99$).
The best linear fit to these data, with an intercept forced to zero, 
gives
\begin{equation}
   A_\text{m} = 78 R_\text{Z} \unit{}{(\micro Hem)} \ . \label{eq_rzam}
\end{equation}
In comparison, the $N_\text{m}$ vs $R_\text{Z}$ data plotted in the 
top left panel of the same figure show a poorer fit, with a linear 
correlation coefficient $r=0.91$ ($r=0.95$ for the corresponding 
13 month running means), with a quadratic curve fitting the data around 
$25\%$ better than a linear one, in terms of rms deviation. 

Focusing on the $A_\text{m}$ vs $R_\text{Z}$ trend, we look at the 
effect of extending the RGO (1876--1976) sequence with USAF--NOAA 
(1977--2013) data.
Superimposed on the upper middle panel of Figure~\ref{f_rzamcor} are the 
13 month running means of the USAF--NOAA $A_\text{m}$ values plotted 
against the 13 month running mean of $R_\text{Z}$ for the same period. 
A linear fit to these uncorrected data gives a slope of 
$\unit{52}{\micro Hem}$, which is precisely $2/3$ of the RGO slope.
This result is in close agreement with \citet[\S~3.2]{Hathaway2010}, who 
indicates that USAF--NOAA's areas should be multiplied by a factor 
$1.48$ in order to match RGO measurement standard.
The top right panel of Figure~\ref{f_rzamcor} shows the 13 month 
running mean of the corrected $A_\text{m}$ plotted against the 13 month 
running mean of $R_\text{Z}$, now with an rms deviation from RGO's 
fit that is more than five times better than for raw USAF--NOAA's 
areas.

The bottom panel of Figure~\ref{f_rzamcor} shows the superimposed 
temporal evolution of $R_\text{Z}$ and $A_\text{m}$, both smoothed and 
unsmoothed, with the correction factor of $1.5$ used for the 1977--2013 
USAF--NOAA sequence.

\subsection{Area distribution}

\begin{figure}
   \plotone{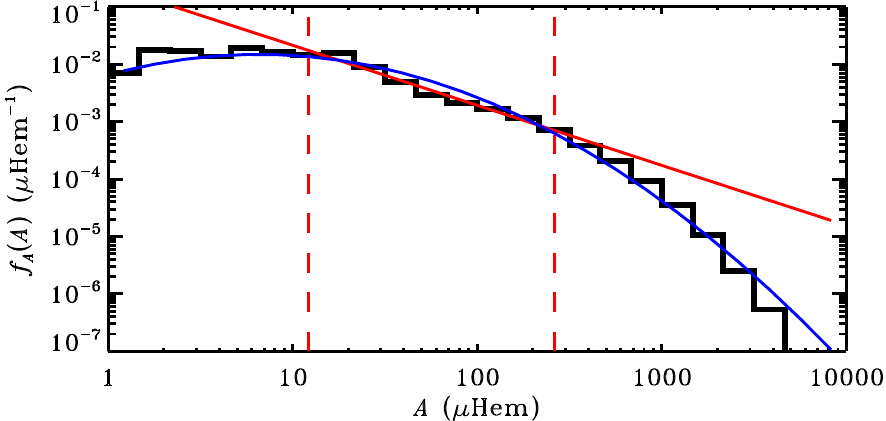}
   \plotone{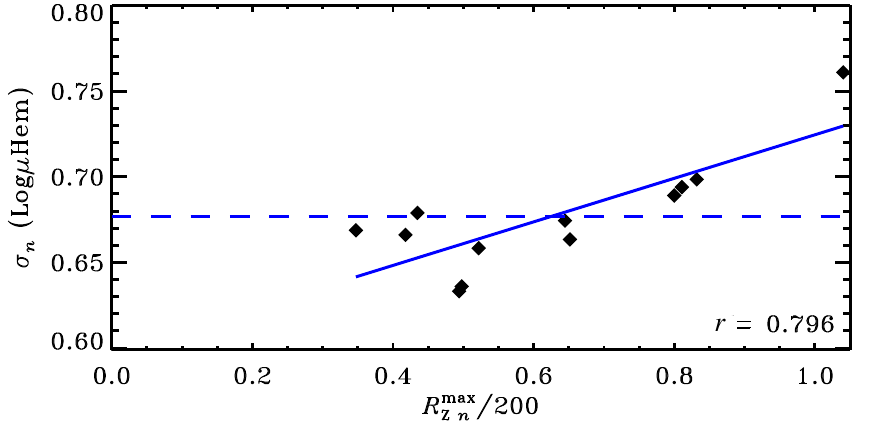}
   \caption{Top panel: {\PDF} of $A$ for all 
   RGO and corrected USAF--NOAA combined, between $1$ and 
   $\unit{6132}{\micro Hem}$, binned every ${\log}A = 0.2$ and properly 
   normalized such that $\int f_A(A) dA = 1$.
   Superimposed is a log-normal best fit, for all bins with more than 
   ten counted sunspot groups (blue line), as well as a fitted power 
   law between $10$ and $\unit{300}{\micro Hem}$ (red line).
   Bottom panel: standard deviations of the 
   log-normal best fits to the {\PDF}s of $A$ for individual cycles, 
   from $n=12$ to $23$, plotted against the corresponding normalized 
   cycle's amplitude $A_{\text{m}~n}^\text{max}$ (black dots). 
   Superimposed are the average of the 12 values (dashed blue line), 
   the linear best fit to those 12 cycles (continuous blue line), as 
   well as the corresponding linear correlation coefficient.}
   \label{f_areapdf}
\end{figure}

We now examine the distribution of areas of RGO and USAF--NOAA sunspot 
groups.
The top panel of Figure~\ref{f_areapdf} illustrates the {\PDF} of $A$, 
computed from regular bins in ${\log}A$.
Following \citet{Bogdan1988}, we carry out a log-normal fit to this 
distribution over the full range of measured areas, which yields a
rms residual three times lower than a power-law fit over the 
restricted interval going from $10$ to $\unit{300}{\micro Hem}$, as 
done in \citet{Jiang2011a}.
Note, however, that in this restricted interval the log-normal fit 
provides a somewhat poorer fit than the power law.
We opt to retain the log-normal fit because it does much better at the
high end of the size spectrum, a proper representation of which is 
critical for the overall surface flux evolution.

We also analyze the {\PDF} of $A$ for individual cycles.
A simultaneous optimization of the mean (${\log}A_0$) and standard 
deviation ($\sigma_{{\log}A}$) of the individual log-normal 
distributions reveals no net tendency for ${\log}A_0$,
neither with respect to cycle amplitude nor length.
However, when fixing ${\log}A_0=1.75$, as obtained from the preceding 
best fit to the whole data set, the standard deviation shows a 
significant dependence on cycle amplitude.
We find a linear correlation $r=0.80$ between $\sigma_{{\log}A}$ and 
the maximum value of the 13 month running mean of $R_\text{Z}$ for each 
cycle ($R_{\text{Z}~n}^\text{max}$).
The best linear fit between those two quantities, as plotted in 
Figure~\ref{f_areapdf} (bottom panel), is roughly three times better 
than the null hypothesis, in terms of rms deviation. Here again, we 
find that a multiplication factor of $1.5$ for USAF--NOAA data is the 
optimal correction for the standard deviation of areas to better 
follow RGO's tendency.
We do not consider any dependence of the area distribution on cycle 
phase or latitude, but this aspect remains to be studied in more 
detail, as suggested in \citet{Jiang2011a}'s analysis. 

In brief, our recipe to produce a set of synthetic emergences begins as 
follows: (i) every month, Equation~(\ref{eq_rzam}) is used to calculate 
the monthly total area $A_\text{m}$ of sunspot groups to emerge; 
(ii) the 
area $A$ of each individual sunspot group is extracted randomly from the 
following log-normal distribution, until $A_\text{m}$ is reached:
\begin{subequations}
\label{eq_areapdf}
\begin{equation}
   f_{A,n}(A) = f_{0,A,n} \frac{1}{A} \exp\biggl({-\frac{
   ({\log}A - {\log}A_0)^2}{2\sigma_{{\log}A,n}^2}}\biggr) \ ,
\end{equation}
with ${\log}A_0=\unit{1.75}{(\log\micro Hem)}$,
\begin{equation}
   \sigma_{{\log}A,n} = 0.60 + 0.13 (R_{\text{Z}~n}^\text{max}/200) 
   \unit{}{(\log\micro Hem)} \ ,
\end{equation}
\end{subequations}
and $f_{0,A,n}$ a normalization factor.

\subsection{Latitudinal distribution and cycle overlaps}

The emergence of sunspots is known to follow the so-called butterfly 
diagram.
Superposing solar cycles 12--20, normalized to the same duration, 
\citet{Jiang2011a} found a quadratic trend for the temporal 
equatorward migration of the average latitude of emergence, with a 
linear increase of average latitude with cycle amplitude.
They also found a quadratic trend for the temporal evolution of 
latitudinal standard deviation around this average latitude.

\begin{figure}
   \plotone{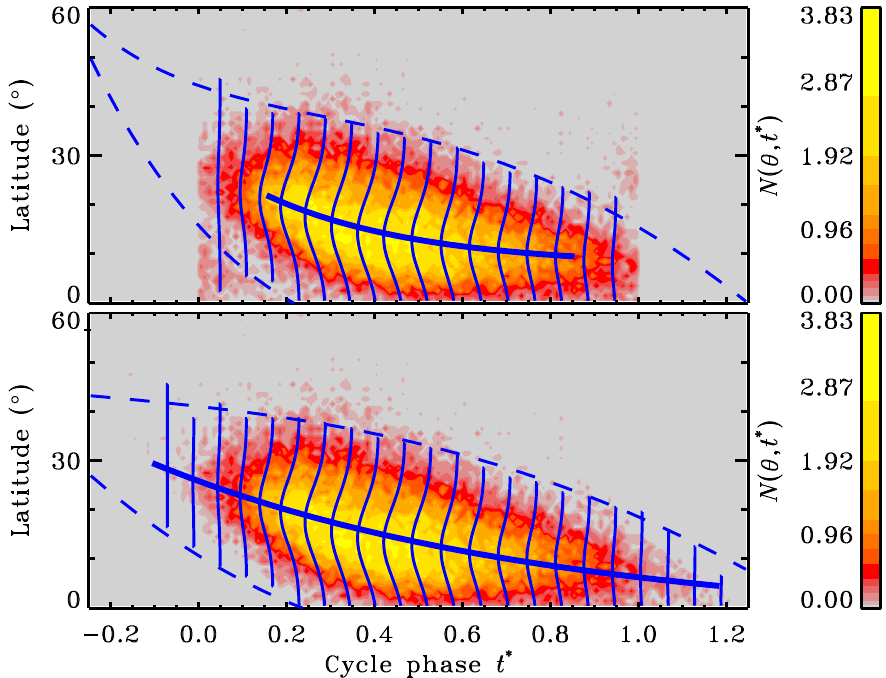}
   \caption{Density plots of the number of emergences inside each 
   time--latitude box, with sample latitudinal gaussian fits 
   (thin blue), 
   exponential trend of the average latitude (thick blue), and 
   $3\sigma_{\lambda}$ lines used to delimit the time--latitude shape of 
   the cycles (dashed blue).
   Top panel: cycles 12 to 23 superimposed according to phase, and 
   averaged ($t^*\in[0.,1.]$).
   Bottom panel: same as upper panel, but with cycle overlaps corrected 
   for and repositioned at the beginning or end of the appropriate 
   cycles (see text).}
   \label{f_butterfly}
\end{figure}

We perform a similar analysis, with cycles 12--23, normalized with 
respect to their respective length $L_n$ and divided into $100$ temporal 
boxes.
Also, instead of considering average latitude and standard 
deviation independently, we directly fit gaussian distributions, inside 
each temporal box, on $1^\circ$ binned latitudinal histograms. 
Figure~\ref{f_butterfly}, upper panel, shows a density plot of the 
number of emergences inside each time--latitude box, averaged over all 
cycles.
Also shown are gaussian fits obtained for a few sample temporal boxes. 
Globally, we find an exponential to fit better the temporal 
equatorward evolution of the average latitude with time than a 
quadratic decrease ($11\%$ improvement in terms of rms deviation). 
Moreover, the standard deviation of the latitudinal spreading increases 
with time until cycle maximum and decreases afterwards, in a roughly 
quadratic manner.

For each cycle, we first perform the fits while avoiding the beginning 
and end of the cycles where overlaps occur.
We use the wing shape delimited by the
$\lambda_{0,n}(t^*) \pm 3\sigma_{\lambda,n}(t^*)$ 
curves, as shown in Figure~\ref{f_butterfly} for all cycles 
superimposed, to characterize overlaps between cycles: every emergence 
inside one cycle's wing shape is assumed to belong to this cycle, while 
emergences outside the region are assigned as an extension of either 
the end of the preceding cycle or the beginning of the following one.
We then calculate again the 
latitudinal gaussian best fits inside each normalized temporal box, the 
best exponential fit on average latitudes, and the best quadratic fit on 
standard deviations, but now for phases $t^*\in[-0.25,1.25]$. We repeat 
the process iteratively until stability is reached.
Figure~\ref{f_butterfly}, bottom panel, illustrates the final curves 
obtained for all cycles superimposed.

The result of this analysis runs as follows:
(i) The probability density of emergence at a given latitude $\lambda$ 
and temporal phase $t^*$ inside each cycle $n$ is given by
\begin{subequations}
\begin{equation}
   f_{\lambda,n}(\lambda,t^*)= f_{0,\lambda,n} \exp\biggl({-\frac{
   (\lambda - \lambda_{0,n}(t^*))^2}{2\sigma_{\lambda,n}(t^*)^2}}\biggr)
   \ ,
\end{equation}
where $\lambda_{0,n}(t^*)$ and $\sigma_{\lambda,n}(t^*)$ are the 
evolving average latitude and standard deviation, and $f_{0,\lambda,n}$ 
a normalization factor.
(ii) The average latitude migrates toward the equator as
\begin{equation}
   \lambda_{0,n}(t^*) = c_{1,n} e^{-t^*/c_{2,n}} + c_{3,n} \unit{}
   {(deg)} \ ,
\end{equation}
%
with an empirical linear dependence of the parameters with respect to 
cycle amplitude: 
$c_{1,n}$ varies from $\unit{19}{\degree}$ for weak cycle 14 to 
$\unit{24}{\degree}$ for strong cycle 19,
$c_{2,n}$ from $0.56$ to $0.65$, and 
$c_{3,n}$ stays roughly constant to $\unit{2.8}{\degree}$.
(iii) The evolution of the latitudinal standard deviation follows a 
quadratic tendency, 
\begin{equation}
   \sigma_{\lambda,n}(t^*) = c_{4,n} + c_{5,n}t^* + c_{6,n}t^{*2} 
   \unit{}{(deg)} \ ,
\end{equation}
\end{subequations}
%
where, again, 
$c_{4,n}$ varies from $\unit{2.1}{\degree}$ to $\unit{3.5}{\degree}$, 
$c_{5,n}$ from $\unit{17.5}{\degree}$ to $\unit{17.4}{\degree}$, and 
$c_{6,n}$ from $\unit{-18.0}{\degree}$ to $\unit{-18.5}{\degree}$.
(iv) For a given cycle, the length of the left overlap with the 
preceding 
cycle ($t^*<0$) is determined by the crossing of the upper and lower 
$3\sigma_{\lambda,n}(t^*)$ curves, while the right overlap ($t^*>1$) 
with the next cycle is determined by the crossing of the upper 
$3\sigma_{\lambda,n}(t^*)$ curve with the equator.
Essentially, this leads to longer overlaps at the beginning of strong 
cycles than at weak cycles.
During the overlapping phase between two cycles, we make the probability 
for an emergence to rise inside the right tail of cycle $n$ to decrease 
linearly from $1$ to $0$, while the probability to rise inside the left 
tail of cycle $n+1$ increases linearly from $0$ to $1$.

The preceding analyses of latitudinal patterns were performed 
simultaneously on the two hemispheres.
Even though hemispheric asymmetries may be self-enhancing through 
sunspot groups nesting 
(see \citealt[\S~4.9]{Hathaway2010}, and references therein), we leave
our synthetic database generator to build such asymmetries solely from 
the stochastic properties of individual sunspot groups and thus set the 
probability to emerge in one hemisphere or the other to $0.5$.

\subsection{Longitudinal distribution}

Sunspot groups have also been found to emerge with a slight preference 
for ``active longitudes'', i.e., near previously emerged sunspot groups
(see \citealt[\S~4.10]{Hathaway2010}, and references therein).
It may be necessary to take this effect into account to properly 
model the extent of open magnetic flux rooted in such active nests.
It is unlikely, though, that a statistical approach would reproduce 
specific realizations of such nesting.
We therefore opt for a uniform random generation of emerging longitudes.

\subsection{Magnetic flux distribution}

We now study the statistical behavior of the WS magnetic database for 
cycle 21, and compare it with the corrected USAF--NOAA area database for 
the same cycle.
While one database uses magnetographic observations, the other is based 
on observations in the visible.
Nevertheless, both present a number of entries of the same order, that 
is respectively $N_\WS = 3047$ {\BMR}s and 
$N_\text{USAF--NOAA} = 3755$ sunspot groups. 
Unfortunately, the independence of the two datasets prevents us from 
performing one-to-one statistics.
Instead, we compare the overall distribution of {\WS} magnetic fluxes 
with the distribution of USAF--NOAA areas.

\begin{figure}
   \plotone{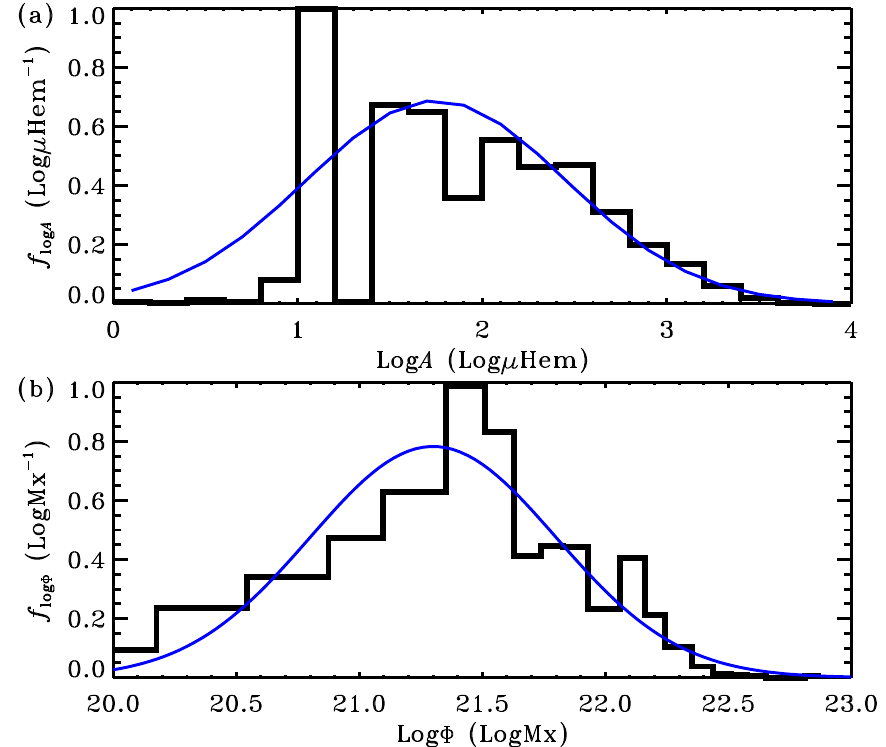}
   \plotone{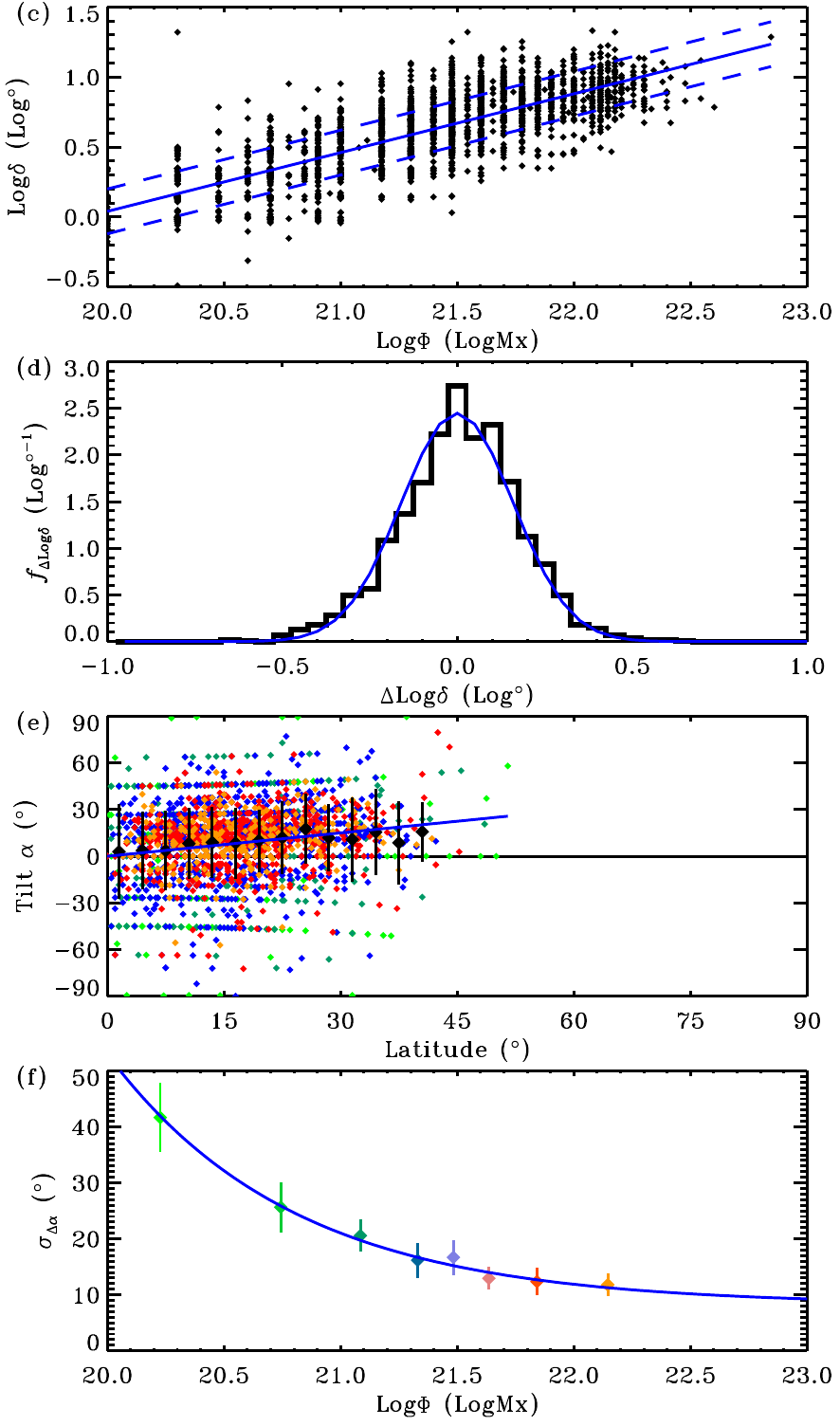}
   \caption{(a) {\PDF} of ${\log}A$ for cycle 21, as extracted from 
   the USAF--NOAA database, with superimposed Gaussian best fit 
   compatible with Equation~(\ref{eq_areapdf}) (blue line).
   (b) {\PDF} of ${\log}\Phi$ as extracted from the {\WS} database, 
   binned irregularly to account for observational biases, with 
   superimposed normal best fit (blue line). 
   (c) Distribution of {\WS} angular separation $\delta$ 
   with respect to their magnetic flux $\Phi$, in log--log scale. 
   Superimposed are a linear best fit (continuous blue line) and 
   $\pm 1 \sigma$ lines (dashed blue).
   (d) {\PDF} of ${\log}\delta$ around the preceding linear fit, with 
   superimposed gaussian best fit (blue line). 
   (e) Distribution of {\WS} tilt angles $\alpha$ against latitude, in 
   eight different colors (green to orange dots) corresponding to eight 
   flux intervals between $10^{20}$ and $\unit{10^{23}}{\maxwell}$. 
   Also shown are the average tilt angle 
   inside $\unit{3}{\degree}$ wide latitudinal bins (black diamonds), 
   the standard deviation inside each latitudinal bin (black lines), and 
   the Joy's linear fit to those averages (blue line).
   (f) Standard deviations, obtained from best gaussian fits of the tilt
   angles spread around the preceding linear fit, as a function
   of magnetic flux, with colors matching those of plot (e). Vertical 
   bars denote relative errors estimated from the rms deviations of 
   individual gaussian fits. Superimposed is the best exponential fit to 
   those points (thin blue line).}
   \label{f_analysewang}
\end{figure}

As for the areas, the {\PDF} of magnetic fluxes ($\Phi$) for cycle 21 
appears roughly log-normal.
Figure~\ref{f_analysewang}b shows the {\PDF} of ${\log}\Phi$ from the 
{\WS} database.
The observed histogram shows a left wing that is slightly too strong, 
but not enough to bring it near a power-law distribution.
We opt to retain the best fit:
\begin{equation}
   f_{\Phi}(\Phi) = f_{0,\Phi} \frac{1}{\Phi} \exp\biggl({-\frac{({\log}\Phi - {\log}\Phi_0)^2}{2\sigma_{{\log}\Phi}^2}}\biggr) \ ,
\end{equation}
with ${\log}\Phi_0 = \unit{21.3}{(\log\maxwell)}$, 
$\sigma_{{\log}\Phi} = \unit{0.5}{(\log\maxwell)}$, and $f_{0,\Phi}$ a 
normalization factor.
As for the size distribution, we do not consider any variability in the 
flux distribution with latitude or with cycle phase, but this aspect
remains to be improved, as suggested in the analysis of 
\citet{Wang1989-0}.

Figure~\ref{f_analysewang}a shows the {\PDF} of ${\log}A$ for cycle 21, 
with a best fit compatible with Equation~(\ref{eq_areapdf}). At first 
glance, the normal distribution of those areas seems questionable. 
However, considering the good fit obtained for all cycles superimposed 
(Figure~\ref{f_areapdf}), and the fact that all USAF--NOAA cycles 
(20--23), but not RGO cycles (12--19), show similar behavior, it is 
more likely that the excess of measurements near ${\log}A = 1$ indicates 
an observational bias that would be responsible for the lack of 
measurements at lower ${\log}A$. This also suggests that the total 
number of sunspot groups measured for cycle 21 could be underestimated. 
As explained earlier, the remaining uncertainty is likely to be of 
minimal 
impact for the purpose of flux transport simulations, since we expect 
the high end of the area and flux spectra to dominate the surface 
evolution.

Assuming a positive correlation between area and magnetic flux, i.e. 
large areas harbor high magnetic fluxes and small areas low magnetic 
fluxes, we can make a direct bridge between the two log-normal 
distributions. With $B_R^\text{mean}$ the average radial surface 
magnetic field inside a given sunspot group, we have
\begin{subequations}
\label{eq_bsmean}
\begin{equation}
   \Phi = B_R^\text{mean} A \ .
\end{equation}
Since the two log-normal distributions do not have the same width, 
$B_R^\text{mean}$ varies with $A$ as follows:
\begin{equation}
   B_R^\text{mean}(A) = \frac{\Phi_0}{A_0} \left(\frac{A}
   {A_0}\right)^{\frac{\sigma_{{\log}\Phi}}{\sigma_{{\log}A,21}}-1} \ ,
\end{equation}
\end{subequations}
with $\Phi_0 = \unit{10^{21.3}}{\maxwell} $, 
$A_0 = \unit{10^{1.75}}{\micro Hem}$, 
$\sigma_{{\log}\Phi} = \unit{0.5}{(\log\maxwell)}$ and 
$\sigma_{{\log}A,21} = \unit{0.70}{(\log\micro Hem)}$, as defined above. 
The result is that small, compact sunspot groups harbor higher average 
magnetic fields 
($B_R^\text{mean}=\unit{3750}{\gauss}$ for $A=\unit{1}{\micro Hem}$) 
than large, spatially extended sunspot groups 
($B_R^\text{mean}=\unit{340}{\gauss}$ for $A=\unit{4000}{\micro Hem}$).
Though this conversion process has obvious limits, we use 
Equation~(\ref{eq_bsmean}) to determine the magnetic flux of every 
synthetic sunspot group generated in Equation~(\ref{eq_areapdf}).
The fact that the area distribution presents a dependence on cycle 
amplitude will thus imply a cycle dependence of the flux distribution as 
well.

We must finally correct the new synthetic database for its absolute 
number of emergences. In fact, while the original numbers of entries in 
{\WS} and USAF--NOAA databases were not equal, the gaussian fits adopted 
worsen the situation: the area under the gaussian curve of 
Figure~\ref{f_analysewang}a now gives 
$N_\text{USAF--NOAA}\simeq4500$ sunspot groups, while that of 
Figure~\ref{f_analysewang}b gives $N_\WS\simeq3000$ {\BMR}s. Assuming 
completeness of the two independent samples, this discrepancy can still 
be justified by the fact that sunspot groups and {\BMR}s are not defined in the  
same manner. To ensure a minimal consistency, we divide the number 
of emergences obtained in the preceding analysis by a factor $1.5$. This 
number will, however, remain a source of uncertainty for determining the 
absolute amount of flux to emerge during a given cycle.

\subsection{Magnetic bipole separations}

Each sunspot group must now be converted into a {\BMR}, that is a pair 
of patches of the same flux and opposite polarity. We consider the 
statistics of angular separation $\delta$ of the bipolar entries in the 
{\WS} database. While there is no obvious trend of $\delta$ with 
latitude, we find a reasonable linear correlation ($r=0.82$) between 
${\log}\delta$ and ${\log}\Phi$. Figure~\ref{f_analysewang}c illustrates 
the repartition of ${\log}\delta$ values with respect to ${\log}\Phi$. 
The linear best fit gives an average value
\begin{equation}
   {\log}\delta_0 = 0.46 + 0.42({\log}\Phi-21) \unit{}{(\log\degree)} 
   \ ,
\end{equation}
with a nearly uniform standard deviation 
$\sigma_{{\log}\delta} = \unit{0.16}{deg}$ around this mean, as 
illustrated by the gaussian fit on 
$\Delta{\log}\delta = {\log}\delta-{\log}\delta_0$ shown in 
Figure~\ref{f_analysewang}d.
A similar analysis can be found in \citet{Wang1989-0}.
Unfortunately, the use of the {\WS} database 
for cycle 21 alone prevents us from looking at any dependence of 
$\delta$ on cycle amplitude.

\subsection{Magnetic bipole tilts}

{\BMR} are known to have their axis tilted with respect to the equator. 
Using again the {\WS} bipolar entries, Figure~\ref{f_analysewang}e plots 
the {\BMR} tilt angles $\alpha$ as a function of latitude. When averaged 
into $\unit{3}{\degree}$ wide latitudinal bins, $\alpha$ shows the 
expected increase with latitude as stated by Joy's law.
We opt for the plain proportional formulation
\begin{subequations}
\begin{equation}
   \alpha_0 = c_\alpha \lambda \ ,
\end{equation}
since other latitudinal profiles used in the literature 
(e.g. $\abs{\alpha_0} \propto \sqrt{\abs{\lambda}}$), 
do not appear to provide any significant improvement. 
We find that values of the proportionality factor $c_\alpha$ varying 
from $0.4$ to $0.6$ could fit the latitudinal trend rather 
similarly, with some unclear dependence on flux.
The value $c_\alpha=0.5$, though, provides the best overall 
compromise when considering the very large, Gaussian-shaped, dispersion 
of $\alpha$ around $\alpha_0$.

On the other hand, standard deviations do show a strong dependence on 
flux amplitude. 
We thus apply gaussian fits to the distribution of 
$\Delta\alpha = \alpha-\alpha_0$ for eight different flux intervals 
between $10^{20}$ and $\unit{10^{23}}{\maxwell}$. 
Figure~\ref{f_analysewang}f illustrates the 
variation of the standard deviation $\sigma_{\Delta\alpha}$ with 
${\log}\Phi$, with relative error bars indicating the importance of the
rms deviation between the gaussian fit and the data in each flux bin. 
We find this standard deviation to decrease exponentially as
\begin{equation}
   \sigma_{\Delta\alpha} = \unit{8.5}{\degree} + \unit{12}{\degree} 
   e^{\textstyle-({\log}\Phi-21)/0.8} \ . \label{eq_sigtilt}
\end{equation}
\end{subequations}
Again, the use of the {\WS} database for cycle 21 prevents us from 
finding any dependence of $\alpha$ on cycle characteristics. 
In a study of observed tilt angles for cycles 15--21, 
\citet{DasiEspuig2010} did find a decrease of average tilt with respect 
to cycle amplitude. 
However, the inclusion of such a result in the construction of our 
synthetic 
database would require a re-evaluation of Equation~(\ref{eq_sigtilt}) 
for cycles other than 21. For simplicity, we chose not to consider 
such systematic dependences of the tilt angles
for the time being.

\bibliographystyle{apj}
\bibliography{references}

\clearpage

\end{document}